\documentclass[onecolumn,showpacs,preprintnumbers]{revtex4}
\usepackage{mathrsfs}
\usepackage{graphicx}
\usepackage{dcolumn}
\usepackage{bm}
\usepackage{epsfig}
\usepackage{amsmath,amssymb}
\usepackage{multirow}

\newcommand{\tabincell}[2]{\begin{tabular}{@{}#1@{}}#2\end{tabular}}
\begin{document}

\title{Systematic analysis of strange single heavy baryons $\Xi_{c}$ and $\Xi_{b}$}
\author{Zhen-Yu Li$^{1}$}
\email{zhenyvli@163.com }
\author{Guo-Liang Yu$^{2}$}
\email{yuguoliang2011@163.com }
\author{Zhi-Gang Wang$^{2}$ }
\email{zgwang@aliyun.com }
\author{Jian-Zhong Gu$^{3}$ }
\author{Jie Lu$^{2}$}
\affiliation{$^1$ School of Physics and Electronic Science, Guizhou Education University, Guiyang 550018,
China\\$^2$ Department of Mathematics and Physics, North China Electric Power University, Baoding 071003,
China\\$^3$ China Institute of Atomic Energy, Beijing 102413,
China}
\date{\today }

\begin{abstract}
Motivated by the experimental progress in the study of heavy baryons, we investigate the mass spectra of strange single heavy baryons in the $\lambda$-mode, where the relativistic quark model and the infinitesimally shifted Gaussian basis function method are employed. It is shown that the experimental data can be well reproduced by the predicted masses. The root mean square radii and radial probability density distributions of the wave functions are analyzed in detail. Meanwhile, the mass spectra allow us to successfully construct the Regge trajectories in the $(J,M^{2})$ plane. We also preliminarily assign quantum numbers to the recently observed baryons, including $\Xi_{c}(3055)$, $\Xi_{c}(3080)$, $\Xi_{c}(2930)$, $\Xi_{c}(2923)$, $\Xi_{c}(2939)$, $\Xi_{c}(2965)$, $\Xi_{c}(2970)$, $\Xi_{c}(3123)$, $\Xi_{b}(6100)$, $\Xi_{b}(6227)$, $\Xi_{b}(6327)$ and $\Xi_{b}(6333)$. At last, the spectral structure of the strange single heavy baryons is shown. Accordingly, we predict several new baryons that might be observed in forthcoming experiments.

\end{abstract}


\maketitle

\section*{I. Introduction}

\label{sec1}
In recent years, many single heavy baryons have been observed in experiments,
 and the mass spectra of single heavy baryon families have become more and more abundant~\cite{art77,art210,art81,art24,art25,art30,art83,art201,art202,art84,art85,art31,art29,art88,art601,art87,art203,art204,art89,art78,art79,art130,art15,art18,art120,art16,art125,art2,art90,art213,art126}. Such a wealth of experimental data gives theorists an opportunity to test the validity of current theoretical frameworks.
Additionally, this is also a good time to carry out a systematic and precise calculation with some theoretical
methods, so as to promote the consistency between the experiments and theories.

The strange single heavy baryon $\Xi_{Q}$ families including $\Xi_{c}$ ($\Xi_{c}^{'}$) ~\cite{art81,art210,art30,art25,art31,art29,art83,art84,art85,art88,art89,art87,art120,art125,art213} and $\Xi_{b}$ ($\Xi_{b}^{'}$)~\cite{art77,art201,art202,art203,art204,art78,art79,art15,art130,art90,art126}, are being established step by step with the cooperative efforts of experimentalists and theorists. So far, more than a dozen $\Xi_{Q}$ baryons have been recorded in the latest particle data group (PDG)~\cite{art2}, even though the $J^{P}$ values of some baryons are still undetermined, such as $\Xi_{c}(3055)$, $\Xi_{c}(3080)$ and $\Xi_{c}(6227)$. Recently, some other $\Xi_{Q}$ baryons have been observed in experiment, including $\Xi_{c}(3123)$~\cite{art31}, $\Xi_{c}(2930)$~\cite{art120}, $\Xi_{c}(2923)$, $\Xi_{c}(2939)$, $\Xi_{c}(2964)$~\cite{art125}, $\Xi_{b}(6327)$ and $\Xi_{b}(6333)$~\cite{art126}.
Accordingly, there have been a lot of theoretical studies on these baryons, such as $\Xi_{c}(3055)$~\cite{art309,art44,art322,art23,art128}, $\Xi_{c}(3080)$~\cite{art23,art45,art70,art311}, $\Xi_{c}(2923)^{0}$ (including $\Xi_{c}(2939)^{0}$ and $\Xi_{c}(2965)^{0}$)~\cite{art321}, $\Xi_{c}(2930)^{0}$~\cite{art32}, $\Xi_{c}(2970)$~\cite{art27,art46,art305,art901,art902,art903}, $\Xi_{c}$(3123)~\cite{art310}, $\Xi_{b}(6227)$~\cite{art319,art318,art313,art306,art316}, $\Xi_{b}(6100)$~\cite{art129,art317} and $\Xi_{b}(6327)$ ($\Xi_{b}(6333)$)~\cite{art320}.
 In order to identify their quantum numbers and assign them suitable positions in
the mass spectra, it is necessary to systematically investigate their spectroscopies.

In recent decades, the heavy baryons have been studied by many theoretical methods, including the quark potential model in the heavy quark-light diquark picture~\cite{art5,art1,art35,art36,art310}, relativistic quark model~\cite{art4}, harmonic oscillator quark model~\cite{art321}, constituent quark model~\cite{art28,art94,art95,art119},  chiral quark model~\cite{art32,art309,art319,art320}, chiral perturbation theory~\cite{art37,art38,art39,art40,art41}, relativistic flux tube model~\cite{art27}, Bethe-Salpeter formalism~\cite{art42}, effective Lagrangian approach~\cite{art318}, $^{3}P_{0}$ decay model~\cite{art33,art34,art43,art47,art48,art49,art50,art76}, lattice QCD~\cite{art51,art52,art53,art54}, bound state picture~\cite{art55}, light cone QCD sum rules~\cite{art56,art57,art58,art59,art60,art61,art62,art63,art64,art65}, and QCD sum rules method~\cite{art66,art67,art68,art69,art71,art72,art73,art74,art75,art308}.

In particular, it is worth mentioning that Ebert $et\ al.$~\cite{art5,art1} put forward a heavy quark-light diquark picture in the framework of a QCD-motivated relativistic quark model, in which an initial three-body problem is reduced to a two-step two-body problem. They systematically studied the spectroscopy and Regge trajectories of heavy baryons, and have achieved a great success in predicting new singly heavy baryons. Because the excited states in the heavy quark-light diquark picture are very similar to those of the $\lambda$-mode in a three-quark system~\cite{art28}, it would be interesting to investigate the $\lambda$-mode in a  three-quark system systematically and study the difference in mass spectra between this mode and the heavy quark-light diquark picture.

 In the 1980s, Godfrey and Isgur developed a relativistic quark model by which they studied
mass spectra of mesons~\cite{art91}. Then, Capstick
and Isgur extended it to the study of baryons~\cite{art4}. In the relativistic quark model, the Hamiltonian contains almost all of the interactions between two quarks, which is expected to give accurate calculations for heavy baryon spectra.

 The Gaussian expansion method (GEM) and the infinitesimally-shifted Gaussian (ISG) basis functions~\cite{art92} have been successfully applied to few-body systems in nuclear physics. ISG method has great advantages to improve the computational accuracy and efficiency in the calculation of few-body systems. In recent years, they were introduced in the study of heavy baryons~\cite{art28,art94,art95}, tetraquarks~\cite{art93,art96,art97} and pentaquarks~\cite{art115}.

  Inspired by the above discussion, we try to combine the relativistic quark model with the ISG method, so as to investigate the strange single heavy baryon spectra in a three-quark system. For the excited states, we would only focus on the $\lambda$-mode and compare the results with those of the heavy quark-light diquark picture and the relevant experimental data as well.
 The present work is a preliminary attempt to systematically investigate strange single heavy baryon spectra and this method is promising to be used in the study of other multi-quark systems, including the exotic ones~\cite{art801,art802,art803,art804,art805}.

This paper is organized as follows. In
Sect.II, we briefly describe the methods used in the theoretical calculations, mainly including the relativistic quark model and the GEM(ISG) method. In Sect.III, we present the root mean square radii and the mass spectra of the $\Xi_{Q}$ baryons, analyze the radial probability distributions, and construct the Regge trajectories. On these bases, we perform a detailed analysis of the baryons that have been of interest recently. At last, the mass spectral structures are displayed.  And Sect.IV is reserved for our conclusions.

\section*{II. Phenomenological methods adopted in this work}

\subsection*{2.1 Relativistic quark model and Jacobi coordinates}

 The relativistic quark model is based on the hypothesis that baryons may be approximately described in terms of center-of-mass (CM) frame valence-quark configurations, the dynamics of which are governed by a Hamiltonian with a one-gluon exchange dominant component at short distances and with a confinement implemented by a flavor-independent Lorentz-scalar interaction~\cite{art4}. For a three-quark system the Hamiltonian reads,
\begin{equation}
\begin{aligned}
&H=H_{0}+V,
\end{aligned}
\end{equation}
\begin{equation}
\begin{aligned}
&H_{0}=\sum_{i=1}^{3}(p_{i}^{2}+m_{i}^{2})^{1/2},\\
\end{aligned}
\end{equation}
\begin{equation}
\begin{aligned}
&V=\sum _{i<j}(\tilde{H}^{conf}_{ij}+\tilde{H}^{so}_{ij}+\tilde{H}^{hyp}_{ij}),
\end{aligned}
\end{equation}
 where $\tilde{H}^{conf}_{ij}$, $\tilde{H}^{so}_{ij}$ and $\tilde{H}^{hyp}_{ij}$ are the confinement, spin-orbit and hyperfine interactions, respectively. The confinement item includes one-gluon exchange potentials and the linear confined potentials. Due to the relativistic effect, the interactions should be modified with CM momentum-dependent factors. It is worth noting that the forms of the interactions in this paper have been rearranged for ease of use~\cite{art116,art93}. The interactions are decomposed as follows:
 \begin{eqnarray}
&\tilde{H}^{conf}_{ij}=G'_{ij}(r)+\tilde{S}_{ij}(r), \\
&\tilde{H}^{so}_{ij}=\tilde{H}^{so(v)}_{ij}+\tilde{H}^{so(s)}_{ij},\\
&\tilde{H}^{hyp}_{ij}=\tilde{H}^{tensor}_{ij}+\tilde{H}^{c}_{ij},
\end{eqnarray}
with
 \begin{eqnarray}
&\tilde{H}^{so(v)}_{ij}=\frac{\textbf{S}_{i}\cdot\textbf{L}_{ij}}{2m^{2}_{i}r_{ij}}\frac{\partial\tilde{G}^{so(v)}_{ii}}{\partial{r_{ij}}}+
\frac{\textbf{S}_{j}\cdot\textbf{L}_{ij}}{2m^{2}_{j}r_{ij}}\frac{\partial\tilde{G}^{so(v)}_{jj}}{\partial{r_{ij}}}+
\frac{(\textbf{S}_{i}+\textbf{S}_{j})\cdot\textbf{L}_{ij}}{m_{i}m_{j}r_{ij}}\frac{\partial\tilde{G}^{so(v)}_{ij}}{\partial{r_{ij}}}, \\
&\tilde{H}^{so(s)}_{ij}=-\frac{\textbf{S}_{i}\cdot\textbf{L}_{ij}}{2m^{2}_{i}r_{ij}}\frac{\partial\tilde{S}^{so(s)}_{ii}}{\partial{r_{ij}}}-
\frac{\textbf{S}_{j}\cdot\textbf{L}_{ij}}{2m^{2}_{j}r_{ij}}\frac{\partial\tilde{S}^{so(s)}_{jj}}{\partial{r_{ij}}}, \\
&\tilde{H}^{tensor}_{ij}=-\frac{\textbf{S}_{i}\cdot\textbf{r}_{ij}\textbf{S}_{j}\cdot\textbf{r}_{ij}/r^{2}_{ij}-\frac{1}{3}\textbf{S}_{i}\cdot\textbf{S}_{j}}{m_{i}m_{j}}
\times(\frac{\partial^{2}}{\partial{r^{2}_{ij}}}-\frac{1}{r_{ij}}\frac{\partial}{\partial{r_{ij}}})\tilde{G}^{t}_{ij}, \\
&\tilde{H}^{c}_{ij}=\frac{2\textbf{S}_{i}\cdot\textbf{S}_{j}}{3m_{i}m_{j}}\nabla^{2}\tilde{G}^{c}_{ij}.
\end{eqnarray}

The modified terms in Eqs. (4), (7), (8), (9) and (10) read,
\begin{eqnarray}
&G'_{ij}=(1+\frac{p^{2}_{ij}}{E_{i}E_{j}})^{\frac{1}{2}}\tilde{G}_{ij}(r_{ij})(1+\frac{p^{2}_{ij}}{E_{i}E_{j}})^{\frac{1}{2}}, \\
&\tilde{G}^{so(v)}_{ij}=(\frac{m_{i}m_{j}}{E_{i}E_{j}})^{\frac{1}{2}+\epsilon_{so(v)}}\tilde{G}_{ij}(r_{ij})(\frac{m_{i}m_{j}}{E_{i}E_{j}})^{\frac{1}{2}+\epsilon_{so(v)}}, \\
&\tilde{S}^{so(s)}_{ii}=(\frac{m_{i}m_{i}}{E_{i}E_{i}})^{\frac{1}{2}+\epsilon_{so(s)}}\tilde{S}_{ij}(r_{ij})(\frac{m_{i}m_{i}}{E_{i}E_{i}})^{\frac{1}{2}+\epsilon_{so(s)}},\\
&\tilde{G}^{t}_{ij}=(\frac{m_{i}m_{j}}{E_{i}E_{j}})^{\frac{1}{2}+\epsilon_{t}}\tilde{G}_{ij}(r_{ij})(\frac{m_{i}m_{j}}{E_{i}E_{j}})^{\frac{1}{2}+\epsilon_{t}},\\
&\tilde{G}^{c}_{ij}=(\frac{m_{i}m_{j}}{E_{i}E_{j}})^{\frac{1}{2}+\epsilon_{c}}\tilde{G}_{ij}(r_{ij})(\frac{m_{i}m_{j}}{E_{i}E_{j}})^{\frac{1}{2}+\epsilon_{c}},
\end{eqnarray}
where $E_{i}=\sqrt{m^{2}_{i}+p^{2}_{ij}}$ is the relativistic kinetic energy, and $p_{ij}$ is the momentum magnitude of either of the
quarks in the CM frame of the $ij$ quark subsystem~\cite{art93}.

$\tilde{G}_{ij}(r_{ij})$ and $\tilde{S}_{ij}(r_{ij})$ are obtained by the smearing transformations of the one-gluon exchange potential $G(r)=-\frac{4\alpha_{s}(r)}{3r}$ and
linear confinement potential $S(r)=br+c$, respectively,
\begin{eqnarray}
\begin{aligned}
&\tilde{G}_{ij}(r_{ij})=\textbf{F}_{i}\cdot\textbf{F}_{j}\sum^{3}_{k=1}\frac{2\alpha_{k}}{\sqrt{\pi}r_{ij}}\int^{\tau_{kij}r_{ij}}_{0}e^{-x^{2}}\mathrm{d}x,
\end{aligned}
\end{eqnarray}
\begin{eqnarray}
\begin{aligned}
&\tilde{S}_{ij}(r_{ij})=-\frac{3}{4}\textbf{F}_{i}\cdot\textbf{F}_{j}\{br_{ij}[\frac{e^{-\sigma^{2}_{ij}r^{2}_{ij}}}{\sqrt{\pi}\sigma_{ij} r_{ij}}+(1+\frac{1}{2\sigma^{2}_{ij}r^{2}_{ij}})\frac{2}{\sqrt{\pi}}\int^{\sigma_{ij}r_{ij}}_{0}e^{-x^{2}}\mathrm{d}x]+c\},
\end{aligned}
\end{eqnarray}
with
\begin{eqnarray}
&\tau_{kij}=\frac{1}{\sqrt{\frac{1}{\sigma^{2}_{ij}}+\frac{1}{\gamma^{2}_{k}}}}, \\
&\sigma_{ij}=\sqrt{s^{2}(\frac{2m_{i}m_{j}}{m_{i}+m_{j}})^{2}+\sigma^{2}_{0}(\frac{1}{2}(\frac{4m_{i}m_{j}}{(m_{i}+m_{j})^{2}})^{4}+\frac{1}{2})}.
\end{eqnarray}
Here $\alpha_{k}$ and $\gamma_{k}$ are constants. $\textbf{F}_{i}\cdot\textbf{F}_{j}$ stands for the inner product of the color matrices of quarks $i$ and $j$. $\textbf{F}$ includes 8 components (the so-called Gell-mann matrices), which can be written as
\begin{eqnarray}
\begin{split}
F_{n}=\left \{
\begin{array}{ll}
   \frac{\hat{\lambda}_{n}}{2},                    & \mathrm{for\ quarks},\\
    -\frac{\hat{\lambda}^{*}_{n}}{2},                    &\mathrm{ for\ antiquarks},
       \end{array}
       \right.
\end{split}
\end{eqnarray}
with $ n=1,\cdot\cdot\cdot,8$. All of the parameters in these formulas are taken from reference~\cite{art4}, except that $b$ and $c$ are revised to be 0.14 GeV$^{2}$ and -0.198 GeV, respectively.

To represent the internal motion of quarks in a few-body system, one commonly introduces the Jacobi coordinates. As shown in Fig.1, there are totally three channels of Jacobi coordinates for the three-body system.
The corresponding Jacobi coordinates are defined as
\begin{eqnarray}
&\boldsymbol\rho_{i}=\textbf{r}_{j}-\textbf{r}_{k}, \\
&\boldsymbol\lambda_{i}=\textbf{r}_{i}-\frac{m_{j}\textbf{r}_{j}+m_{k}\textbf{r}_{k}}{m_{j}+m_{k}},
\end{eqnarray}
where $i$, $j$, $k$ = 1, 2, 3(or replace their positions in turn). $\textbf{r}_{i}$ and $m_{i}$ denote the position vector and the mass of the $i$th quark, respectively.

\begin{figure}[htbp]
\begin{center}
\includegraphics[width=0.8\textwidth]{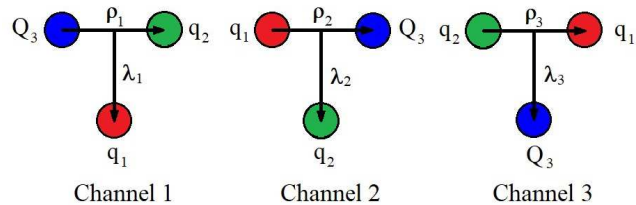}
\end{center}
\caption{(Color online)Jacobi coordinates for the three-quark system. We denote the heavy quark as the 3rd particle in the case of single heavy baryons. }
\end{figure}

 We perform the calculations based on the channel 3. In this case, the 3rd quark is just the heavy quark, which is consistent with the heavy quark limit~\cite{art118,art117}. What is more, $\textbf{\emph{l}}_{\rho3}$ (denoted in short as $\textbf{\emph{l}}_{\rho}$) is clearly defined as the orbital angular momentum between the light quarks, and  $\textbf{\emph{l}}_{\lambda3}$ (denoted in short as $\textbf{\emph{l}}_{\lambda}$) represents the one between the heavy quark and the light-quark pair.

\subsection*{2.2 The heavy quark limit and wave function }

In the heavy quark limit, the heavy quark within the heavy baryon system is decoupled from the
two light quarks. With the requirement of the flavor SU(3) subgroups for the light quarks, the baryons belong to either a sextet $(6_{F})$ of flavor symmetric states $\Xi_{Q}^{'}$, or an antitriplet $(\bar{3}_{F})$ of the flavor antisymmetric states $\Xi_{Q}$.
The flavor wave functions of strange single heavy baryons are written as,
\begin{eqnarray}
\begin{aligned}
&& \Xi_{Q}^{'}=\frac{1}{\sqrt{2}}(qq_{s}+q_{s}q)Q, \\
&& \Xi_{Q}=\frac{1}{\sqrt{2}}(qq_{s}-q_{s}q)Q.
\end{aligned}
\end{eqnarray}
Here $q$ denotes up or down quark, and $Q$ is charm or bottom quark. $q_{s}$ is strange quark. For a quantum state with specified angular momenta in this work, the spatial wave function is combined with the spin function as follow.
 \begin{eqnarray}
\begin{aligned}
|l_{\rho} \ l_{\lambda} \ L \ s\ j\ J\ M_{J}\rangle
&=\sum^{1/2}_{m_{s3}=-1/2}\sum^{j}_{m_{j}=-j}\sum^{L}_{M_{L}=-L}\sum^{s}_{m_{s}=-s}\sum^{1/2}_{m_{s1}=-1/2}\sum^{1/2}_{m_{s2}=-1/2}
\sum^{l_{\rho}}_{m_{\rho}=-l_{\rho}}\sum^{l_{\lambda}}_{m_{\lambda}=-l_{\lambda}}\\
&\ \times\langle j\ m_{j}\ s_{3}\ m_{s_{3}}| j\ s_{3}\ J\ M_{J}\rangle\times \langle L\ M_{L}\ s\ m_{s}| L\ s\ j\ m_{j}\rangle\\
&\times\langle s_{1}\ m_{s_{1}}\ s_{2}\ m_{s_{2}}| s_{1}\ s_{2}\ s\ m_{s}\rangle\times \langle l_{\rho}\ m_{\rho} l_{\lambda}\ m_{\lambda}| l_{\rho}\ l_{\lambda}\ L\ M_{L}\rangle\\
&\times |l_{\rho}\ m_{\rho} \rangle\otimes |l_{\lambda}\ m_{\lambda} \rangle\otimes |s_{1}\ m_{s_{1}} \rangle\otimes
 |s_{2}\ m_{s_{2}} \rangle \otimes |s_{3}\ m_{s_{3}} \rangle ,
\end{aligned}
\end{eqnarray}
with $\textbf{L}=\textbf{\emph{l}}_{\rho}+\textbf{\emph{l}}_{\lambda}$, $\textbf{s}=\textbf{s}_{1}+\textbf{s}_{2}$, $\textbf{j}=\textbf{L}+\textbf{s}$,  $\textbf{J}=\textbf{j}+\textbf{s}_{3}$.
 $l_{\rho}$, $l_{\lambda}$, $L$, $s$, $j$, $J$ and $M_{J}$  are the quantum numbers which characterize a given quantum state in theory. This scheme $|l_{\rho} \ l_{\lambda} \ L \ s\ j\ J\ M_{J}\rangle$ ($j$-$s$ coupling) is also commonly used to analyze the strong decay of heavy baryons~\cite{art48}.

\subsection*{2.3 GEM and ISG}

In calculations, the spatial wave function $|l_{\rho}\ m_{\rho} \rangle\otimes |l_{\lambda}\ m_{\lambda} \rangle$ in formula (24) should be expanded in a set of basis functions. Naturally, one of the candidates is the simple harmonic oscillator(SHO) basis for its good orthogonality. However, the completeness of the SHO is not rigorous in calculations because a truncated set has to be used~\cite{art91,art4}. Compared to the SHO basis functions, the advantage of the Gaussian basis functions is that they can form an approximately complete set in a finite coordinate space.

Following formula (24), the  spatial wave function is expanded in terms of a set of Gaussian basis functions,
 \begin{eqnarray}
\begin{aligned}
 |l_{\rho} m_{\rho} \rangle\otimes |l_{\lambda} m_{\lambda} \rangle=\sum_{n_{\rho}=1}^{n_{max}}\sum_{n_{\lambda}=1}^{n_{max}}c_{n_{\rho}n_{\lambda}}|n_{\rho} l_{\rho} m_{\rho} \rangle^{G} \otimes |n_{\lambda} l_{\lambda} m_{\lambda} \rangle^{G},
\end{aligned}
\end{eqnarray}
where the Gaussian basis function $|nlm \rangle^{G}$ is commonly written in  position space as
 \begin{eqnarray}
\begin{aligned}
 &\phi^{G}_{nlm}(\textbf{r})=\phi^{G}_{nl}(r)Y_{lm}(\hat{\textbf{r}}),\\
 &\phi^{G}_{nl}(r)=N_{nl}r^{l}e^{-\nu_{n}r^{2}},\\
 &N_{nl}=\sqrt{\frac{2^{l+2}(2\nu_{n})^{l+3/2}}{\sqrt{\pi}(2l+1)!!}},
\end{aligned}
\end{eqnarray}
or in momentum space as
\begin{eqnarray}
\begin{aligned}
 &\phi^{'G}_{nlm}(\textbf{p})=\phi^{'G}_{nl}(p)Y_{lm}(\hat{\textbf{p}}),\\
 &\phi^{'G}_{nl}(p)=N^{'}_{nl}p^{l}e^{-\frac{p^{2}}{4\nu_{n}}},\\
 &N^{'}_{nl}=(-i)^{l}\sqrt{\frac{2^{l+2}}{\sqrt{\pi}(2\nu_{n})^{l+3/2}(2l+1)!!}},
\end{aligned}
\end{eqnarray}
with
\begin{eqnarray}
\begin{aligned}
& \nu_{n}=\frac{1}{r^{2}_{n}},\\
& r_{n}=r_{1}a^{n-1}\ \ \ (n=1,\ 2,\ ...,\ n_{max}).
\end{aligned}
\end{eqnarray}
$r_{1}$, $a$, and $n_{max}$ are the Gaussian size
parameters in the geometric progression for numerical calculations, and the final results are stable and independent of
these parameters within an approximately complete set in a sufficiently large space.

The Gaussian basis functions are non-orthogonal, which leads to a generalized matrix eigenvalue problem,
\begin{equation}
\begin{aligned}
 \sum_{\kappa,\kappa'=1}^{\kappa_{max}}[H_{\kappa\kappa'}-E\tilde{N}_{\kappa\kappa'}]c_{\kappa'}=0,
\end{aligned}
\end{equation}
with
\begin{equation}
\begin{aligned}
 \tilde{N}_{\kappa\kappa'}=\langle\phi^{G}_{n_{\rho}l_{\rho}m_{\rho}}|\phi^{G}_{n_{\rho'}l_{\rho'}m_{\rho'}}\rangle
 \times\langle\phi^{G}_{n_{\lambda}l_{\lambda}m_{\lambda}}|\phi^{G}_{n_{\lambda'}l_{\lambda'}m_{\lambda'}}\rangle \\
 =(\frac{2\sqrt{\nu_{n_{\rho}}\nu_{n_{\rho'}}}}{\nu_{n_{\rho}}+\nu_{n_{\rho'}}})^{l_{\rho}+3/2}
 \times(\frac{2\sqrt{\nu_{n_{\lambda}}\nu_{n_{\lambda'}}}}{\nu_{n_{\lambda}}+\nu_{n_{\lambda'}}})^{l_{\lambda}+3/2},
\end{aligned}
\end{equation}
where $\kappa=1,\ 2,\ ...,\ \kappa_{max}$, $\kappa_{max}=n_{max}\times n_{max}$ and $c_{\kappa}=c_{n_{\rho}n_{\lambda}}$. $H$ and $E$ denote the Hamiltonian and the eigenvalue, respectively.

 In the calculation of Hamiltonian matrix elements of three-body systems, particularly when
complicated interactions are employed, integrations over all of the radial and angular coordinates
become laborious even with the Gaussian basis functions. This process can be simplified by introducing
the ISG basis functions~\cite{art92}.

\section*{III. Numerical results and discussions}

\subsection*{3.1 Numerical stabilities and the $\lambda$-mode  }

 To obtain stable numerical solutions, the Gaussian size
parameters set $\{n_{max},\ r_{1},\ r_{n_{max}}\}$ should be optimized. For the Gaussian functions which are a set of non-orthogonal bases in a finite coordinate space, the number of the bases should be in a reasonable range. As shown in Fig.2, the numerical stability is achieved when the dimension parameter $n_{max}$ falls in the range of $9\sim14$, with  $r_{1}=0.18$GeV$^{-1}$ and $r_{n_{max}}=15$GeV$^{-1}$. $n_{max}=10$ is finally adopted in this work, with which both the computation efficiency and accuracy are actually satisfied.
\begin{figure}[htbp]
\begin{center}
\includegraphics[width=0.4\textwidth]{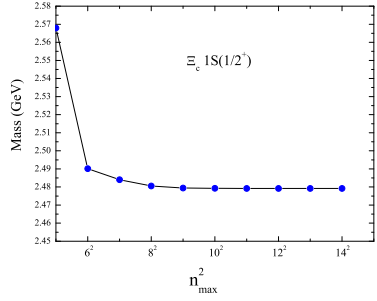}
\end{center}
\caption{(Color online) Numerical stability of $\Xi_{c}1S(\frac{1}{2}^{+})$ mass with respect to the dimension parameter $n_{max}$.}
\end{figure}

  $nL(J^{P})$ is commonly used to describe a baryon state. If angular momentum $L\neq0$, there exist several $|l_{\rho} l_{\lambda}  L  s j J M_{J}\rangle$ states under the condition  $\textbf{L}=\textbf{\emph{l}}_{\rho}+\textbf{\emph{l}}_{\lambda}$. They may be divided into the following three modes: (1) The $\rho$-mode with $l_{\rho}\neq0$ and $l_{\lambda}=0$; (2) The $\lambda$-mode with $l_{\rho}=0$ and $l_{\lambda}\neq0$; (3) The $\lambda$-$\rho$ mixing mode with $l_{\rho}\neq0$ and $l_{\lambda}\neq0$.

 As an example, the excitation energies of the $1P(\frac{1}{2}^{-}, \frac{3}{2}^{-})_{j=1}$  states of $\bar{3}_{F}$ as functions of $m_{Q}$ are investigated, where the dependence of excitation energies on $m_{Q}$ of the $\lambda$-mode is compared with that of the $\rho$-mode. As shown in Fig.3, the $\lambda$-mode and the $\rho$-mode are clearly separated when $m_{Q}$ increases from 1.0 GeV to 5.0 GeV. In the case of $6_{F}$, we come to the same conclusion when $m_{Q}>1.5$GeV as shown in Fig.4.
Besides of $p$-wave states, the same feature is also shown in higher angular excited states actually.
 This conclusion has been obtained in reference~\cite{art28}. Thus, we investigate the mass spectrum in the $\lambda$-mode.

\begin{figure}[h]
\begin{minipage}[h]{0.45\linewidth}
\centering
\includegraphics[height=5cm,width=6cm]{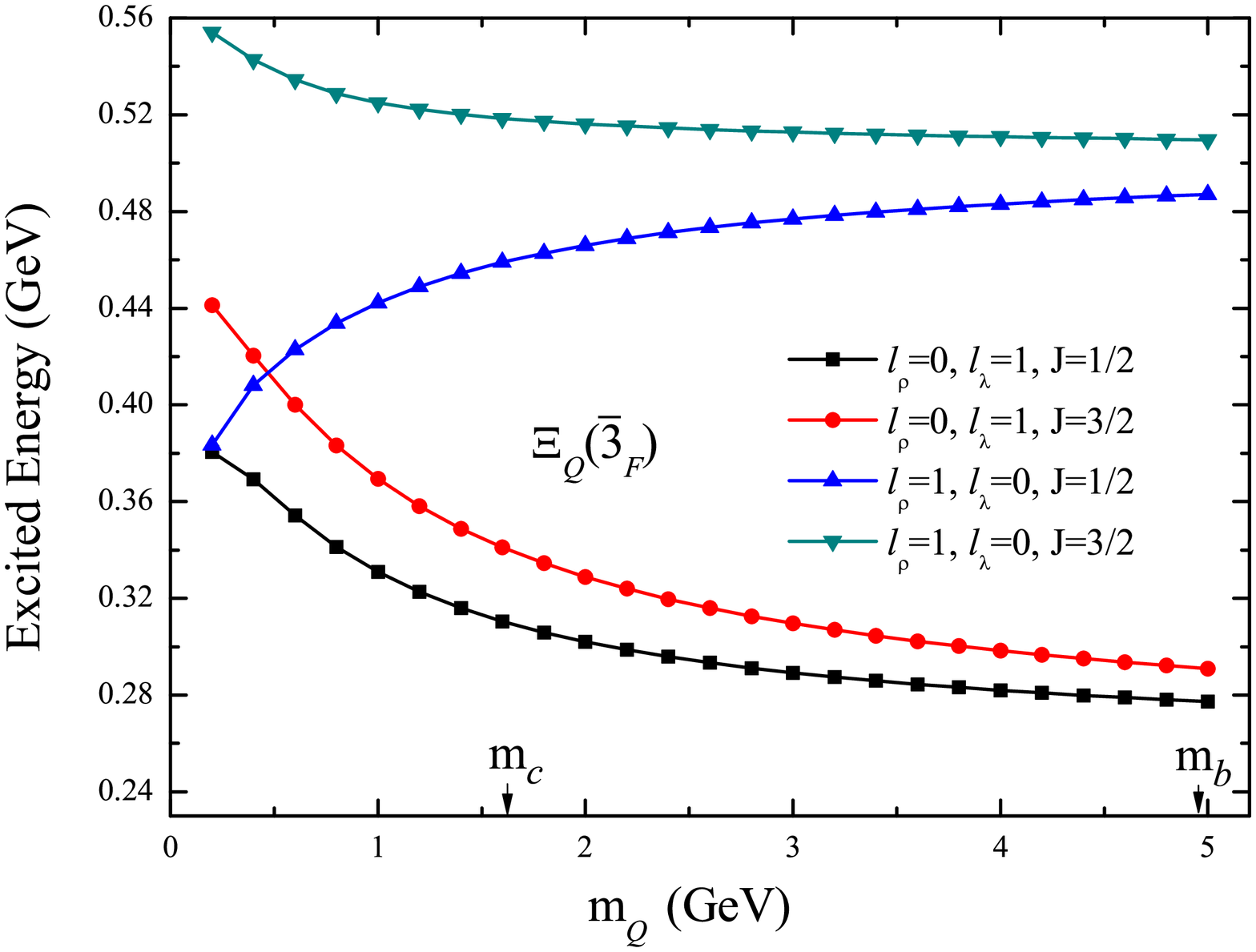}
\caption{(Color online)The dependence of excitation energies on $m_{Q}$ for different modes of $\Xi_{Q}$. The black and red curves represent the $\lambda$-mode. The blue and green ones denote the $\rho$-mode. }
\end{minipage}
\hfill
\begin{minipage}[h]{0.45\linewidth}
\centering
\includegraphics[height=5cm,width=6cm]{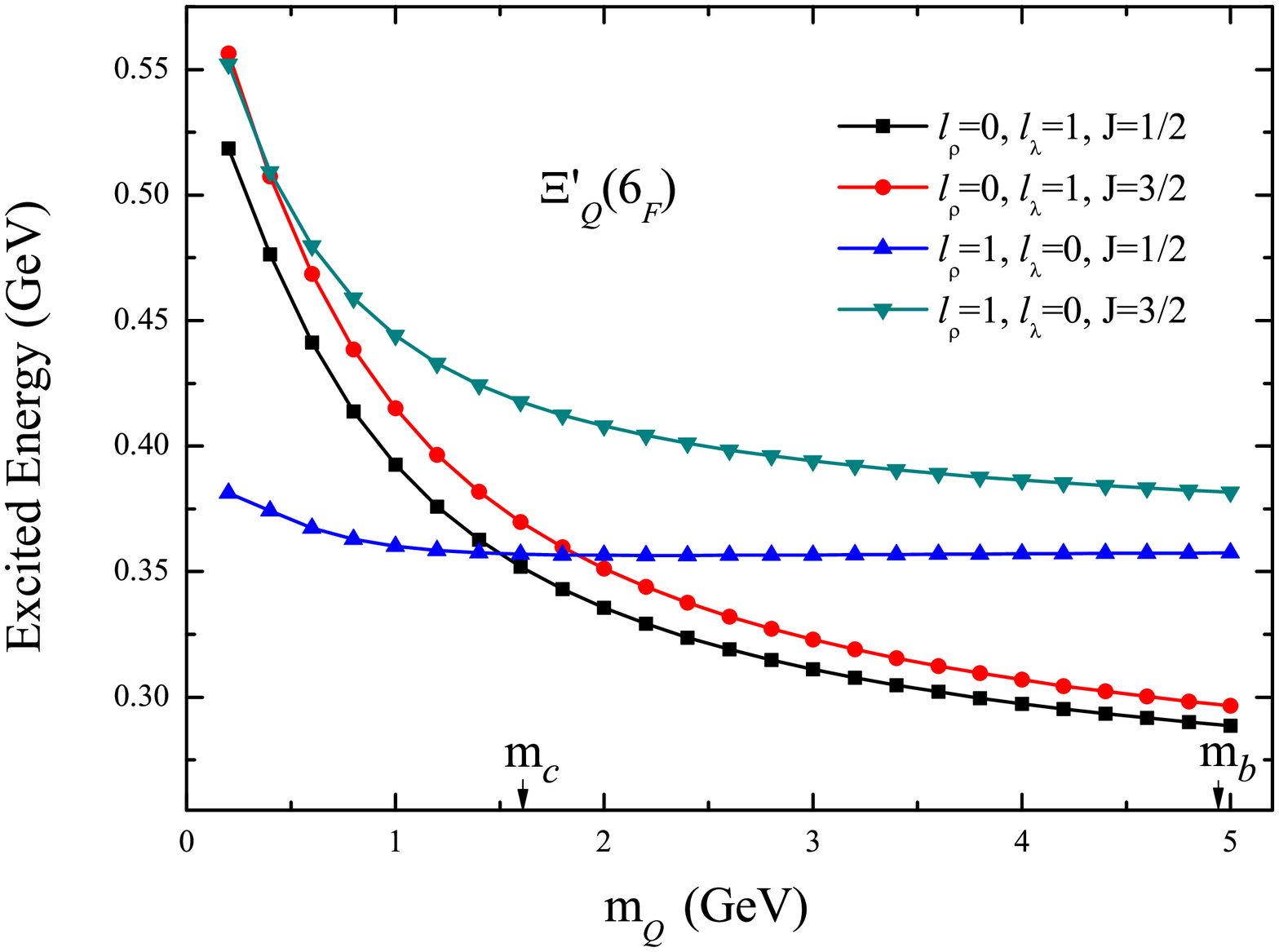}
\caption{(Color online)The excitation energies versus $m_{Q}$ for different modes of $\Xi_{Q}^{'}$. Note that the  black curve is lower than the blue one with $m_{Q}>1.5$GeV and the red curve is overall lower than the green one ($m_{c}=1.628$ GeV and $m_{b}=4.977$ GeV are taken in this work).}
\end{minipage}
\end{figure}

\subsection*{3.2 Mass spectra, root mean square radius and radial probability density distribution }

In this subsection, the root mean square radii, radial probability density distributions and the mass spectra of strange single heavy baryons are presented. For convenience, the relevant experimental data are given together. The detailed results are listed in Tables I-VI (see the appendix). There are a total of four families, namely $\Xi_{c}$, $\Xi_{c}^{'}$, $\Xi_{b}$ and $\Xi_{b}^{'}$. The mass spectra of excited states with quantum numbers up to $n=4$ and $L=4$ are displayed. There have been a lot of other theoretical works on this subject~\cite{art23,art70,art411,art12,art318}. Among them, Ebert $et\ al.$ have studied the heavy baryon spectra with a quark potential model in the heavy quark-light diquark picture~\cite{art1}. As an important reference, their numerical results are also placed in the tables.

 Through the analysis of these calculated results, some general features of the mass spectra are summarized as follows:
  First, $\Xi_{Q}$  is lower than $\Xi_{Q}^{'}$ in energy. This feature has been recognized in light
baryons where the highly orbitally excited states have an antisymmetric structure which minimizes the energy~\cite{art122};
  Second, the mass splitting of spin-doublet states becomes smaller with increasing $L$. For example, Table I shows the mass differences of the spin-doublets of $1P$-, $1D$-, $1F$-, and $1G$-wave are 30 MeV, 13 MeV, 5 MeV, and 1 MeV, respectively;
  Thirdly, for the same $L$, the mass splitting  hardly changes with the increase of $j$. For example, Tables II and III show  the mass differences of $1D$ doublets with $j=1,2,3$ are 10 MeV, 10 MeV and 13 MeV, respectively;
  The last, the mass difference between the two adjacent radial excited states gradually decreases with increasing $n$, which is clearly different from that given by Ebert $et\ al.$.

On the other hand, the calculated root mean square radii and radial probability density distributions carry important information. For a three-quark system, the radial probability densities $\omega(r_{\rho})$ and $\omega(r_{\lambda})$ can be defined as follows,
\begin{eqnarray}
\begin{aligned}
& \omega(r_{\rho})=\int |\Psi(\textbf{r}_{\rho},\textbf{r}_{\lambda})|^{2}\mathrm{d}\textbf{r}_{\lambda}\mathrm{d}\Omega_{\rho},\\
&  \omega(r_{\lambda})=\int |\Psi(\textbf{r}_{\rho},\textbf{r}_{\lambda})|^{2}\mathrm{d}\textbf{r}_{\rho}\mathrm{d}\Omega_{\lambda},
\end{aligned}
\end{eqnarray}
where $\Omega_{\rho}$ and $\Omega_{\lambda}$ are the solid angles spanned by vectors $\textbf{r}_{\rho}$ and $\textbf{r}_{\lambda}$, respectively. From Figs.5-7 and tables I-VI, one can find some interesting properties.

(1) For the same $n$ states, when $L$ changes from 1 to 4, their $\langle r_{\rho}^{2}\rangle ^{1/2}$ values increase small. But their $\langle r_{\lambda}^{2}\rangle ^{1/2}$ values become larger gradually. The similar phenomenon can be seen in Figs.5 and 6, where the radial probability of $r_{\rho}^{2}\omega(r_{\rho})$ changes a little with different $L$ values. However, the peak of the $r_{\lambda}^{2}\omega(r_{\lambda})$ significantly shifts outward with increasing $L$.

(2) For the same $L$ states, $\langle r_{\rho}^{2}\rangle ^{1/2}$ and $\langle r_{\lambda}^{2}\rangle ^{1/2}$ generally become larger with increasing $n$. And the peaks of their probability densities are generally shifted outward, as shown in Fig.7.

(3) The shapes of the eight black (solid) lines in Fig.5 are almost as same as those in Fig.6. And the values of $\langle r_{\rho}^{2}\rangle ^{1/2}$ for the same state are almost the same for $\Xi_{c}$($\Xi_{c}^{'}$) and $\Xi_{b}$($\Xi_{b}^{'}$) families. This reflects the fact that the configurations of the two light quarks in $\Xi_{c}$($\Xi_{c}^{'}$) and $\Xi_{b}$($\Xi_{b}^{'}$) baryons are similar to each other.

(4) As shown in tables I-VI, the root mean square radii of those baryons which have been experimentally well established are generally less than 0.8 fm.

 When the root mean square radius becomes larger, the radial probability distribution of the wave function appears more outwardly extended, and the baryons become even looser. Generally speaking, the root mean square radius of a compact baryon might be within a threshold, which is of help to estimate the upper limit of the mass spectrum and constrain the number of members for each heavy baryon family.

\begin{figure}[htbp]
\begin{center}
\includegraphics[width=1.0\textwidth]{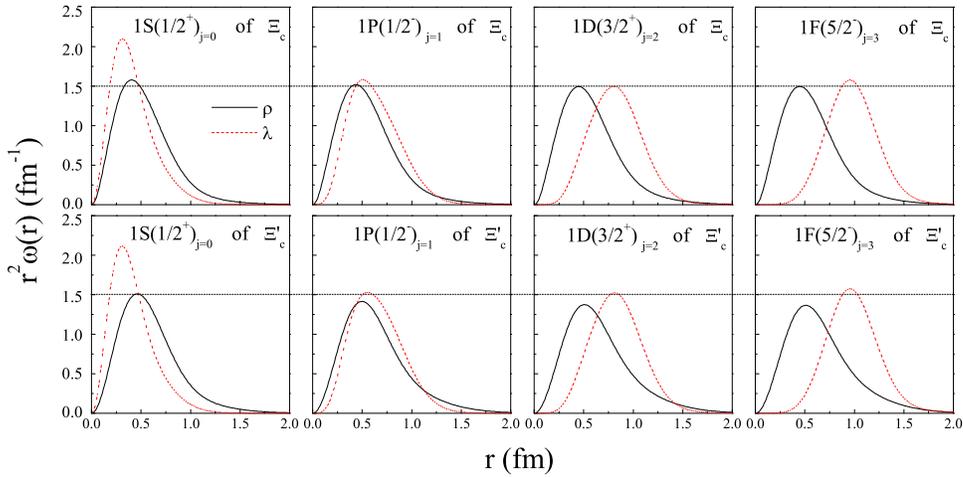}
\end{center}
\caption{(Color online)Radial probability density distributions for some $1L$ states in the $\Xi_{c}$ and $\Xi_{c}^{'}$ families. The solid line denotes the probability density with $r_{\rho}$, and the dash line denotes the one with $r_{\lambda}$. }
\end{figure}

\begin{figure}[htbp]
\begin{center}
\includegraphics[width=1.0\textwidth]{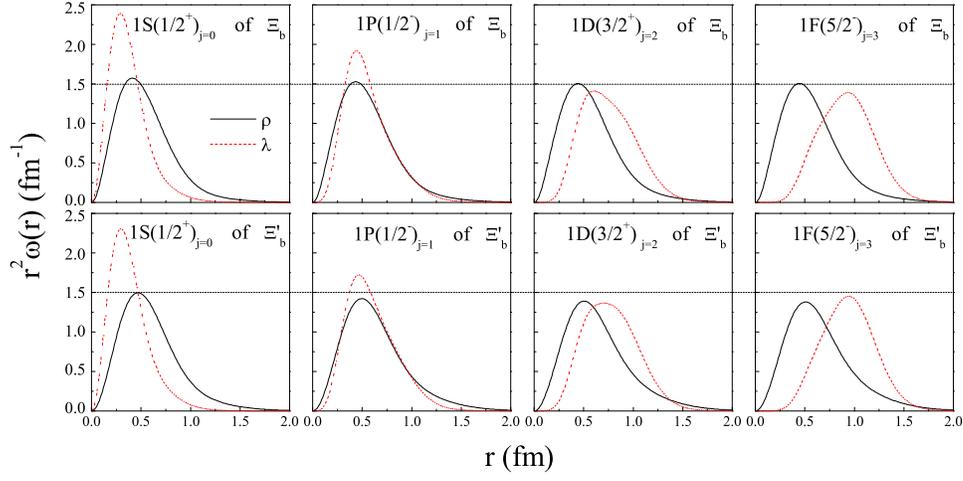}
\end{center}
\caption{(Color online)Same as Fig.5, but for the $\Xi_{b}$ and $\Xi_{b}^{'}$ families.}
\end{figure}

\begin{figure}[htbp]
\begin{center}
\includegraphics[width=0.8\textwidth]{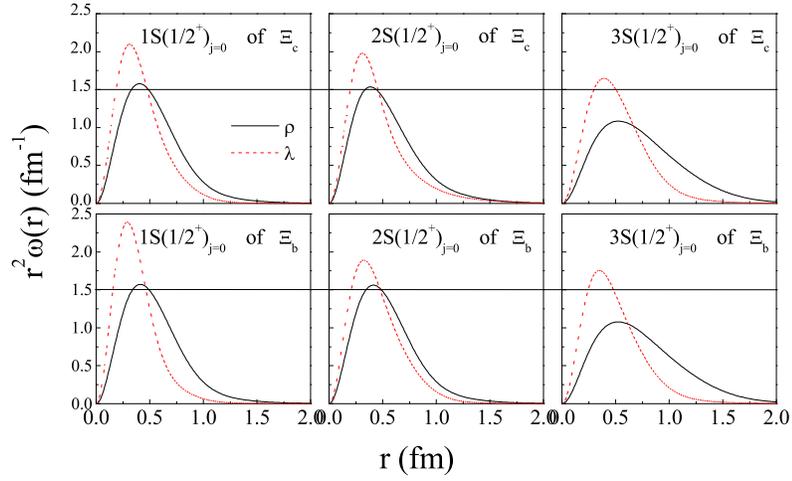}
\end{center}
\caption{(Color online)Radial probability density distributions for some $nS$ states in the $\Xi_{c}$ and $\Xi_{b}$ families.}
\end{figure}

\subsection*{3.3 Regge trajectories}

 Regge trajectories is an effective method to discribe the hadron mass spectrum\cite{art503,art504,art501,art502,art505}.
  In 2011, Ebert $et\ al.$ constructed the heavy baryon Regge trajectories in both the $(J, M^{2})$ and the $(n, M^{2})$ planes~\cite{art1}.

In this subsection, we investigate the Regge trajectories in the $(J, M^{2})$ plane based on our calculated mass spectra. The states in a baryon family can be classified according to the following parities and angular momenta: (1) Natural $P=(-1)^{J+1/2}$ and unnatural $P=(-1)^{J-1/2}$ parities (written in short as $NP$ and $UP$, respectively)\cite{art124}; (2) $J=j+1/2$ and $J=j-1/2$ (written in short as $NJ$ and $UJ$). Thus, the states in the $\Xi_{c}$ or $\Xi_{b}$ family are divided into two groups, and the states in the $\Xi_{c}^{'}$ or $\Xi_{b}^{'}$ family are divided into six groups.
In this paper, we use the following definition for the $(J, M^{2})$ Regge trajectories,
 \begin{eqnarray}
M^{2}=\alpha J+ \beta,
\end{eqnarray}
where $\alpha$ and $\beta$ are the slope and intercept. In Figs. 7 and 8, we plot the Regge trajectories in the $(J, M^{2})$
plane. The three lines in each figure correspond to the radial quantum number $n$= 1, 2, 3, respectively.
 The fitted slopes and intercepts of the Regge trajectories are given in Tables VII and VIII.

It is shown that the linear trajectories appear clearly in the $(J, M^{2})$ plane. All the data points fall on the trajectory lines. This indicates that the Regge trajectory has a strong universality and our theoretical calculations are reliable. These trajectories are almost parallel, but not equidistant, which is an apparent difference between our mass spectra and those in reference~\cite{art1}.

In this paper, we do not show the Regge trajectories in the $(n, M^{2})$ plane. In fact, the linear trajectories in the $(n, M^{2})$ plane can not be constructed from our predicted masses. As will be mentioned in subsection 3.5, if the observation of heavy baryons in forthcoming experiments touched the $3S$ sub shell, it would be a chance to check the $(n, M^{2})$ Regge trajectories, and then we can determine whether a single heavy baryons is a three-quark system or a quark-diquark system.

\begin{figure}[htbp]
\begin{center}
\includegraphics[width=1.0\textwidth]{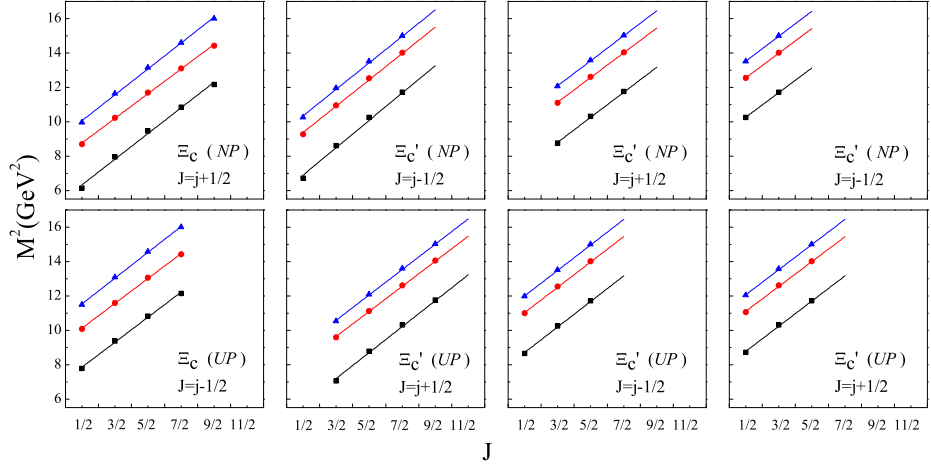}
\end{center}
\caption{(Color online)$(J, M^{2})$ Regge trajectories for the $\Xi_{c}$  ($\Xi_{c}^{'}$) families and $M^{2}$ is in GeV$^{2}$.}
\end{figure}

\begin{figure}[htbp]
\begin{center}
\includegraphics[width=1.0\textwidth]{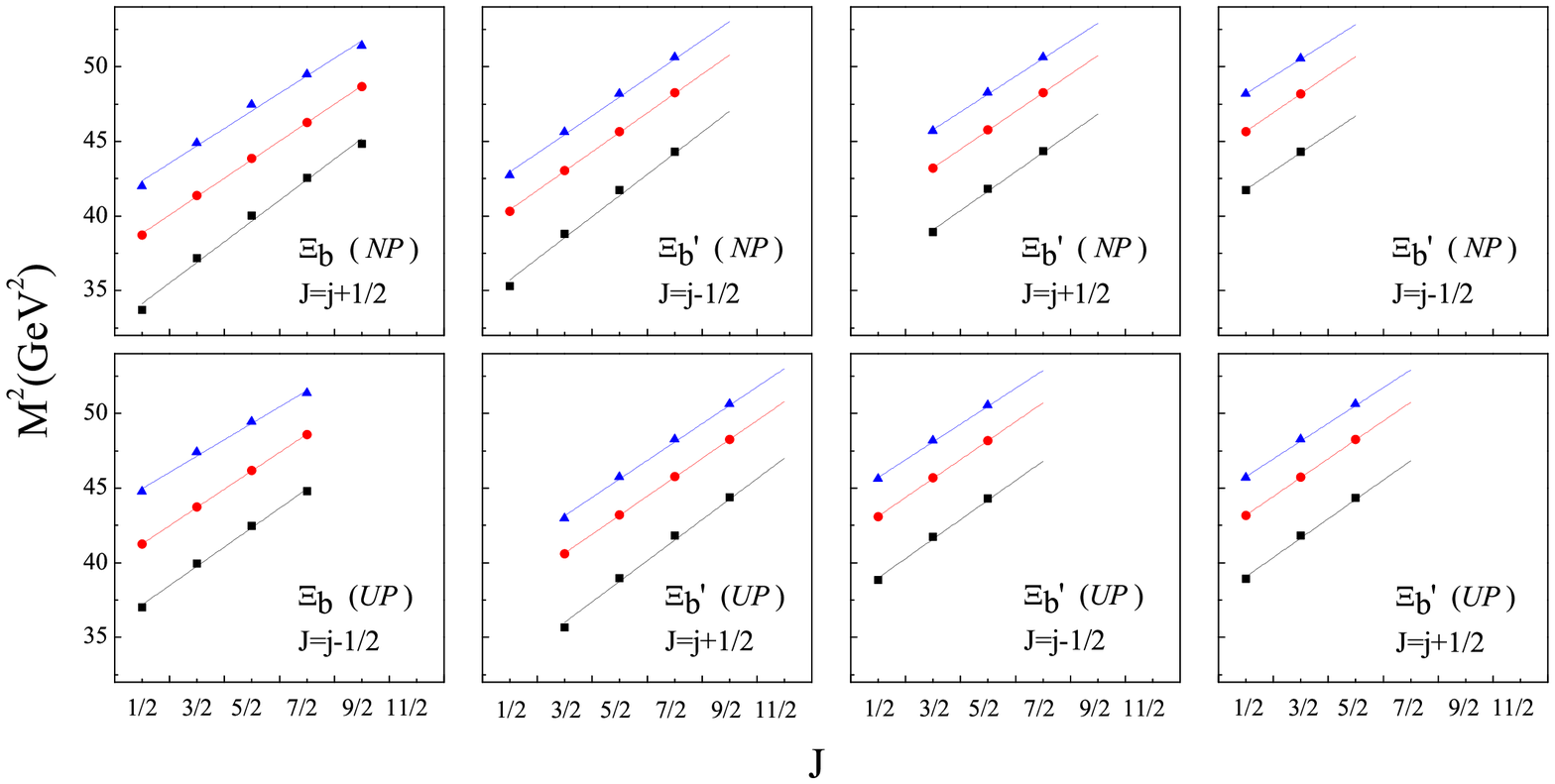}
\end{center}
\caption{(Color online)$(J, M^{2})$ Regge trajectories for the $\Xi_{b}$ ($\Xi_{b}^{'}$) families and $M^{2}$ is in GeV$^{2}$.}
\end{figure}

\subsection*{3.4 Preliminary assignment to some observed heavy baryons }
For the well determined $\Xi_{c}$ ($\Xi^{'}_{c}$) baryons in the PDG, we can assign them to the corresponding positions nicely as follows, $\Xi_{c}^{+}$ and $\Xi_{c}^{0}$ $\leftrightarrow$ $\Xi_{c}$ $1S(\frac{1}{2}^{+})$, $\Xi_{c}^{'+,0}$ $\leftrightarrow$ $\Xi_{c}$ $1S(\frac{1}{2}^{+})$, $\Xi_{c}(2645)^{+,0}$ $\leftrightarrow$  $\Xi_{c}^{'}$ $1S(\frac{3}{2}^{+})$, $\Xi_{c}(2790)^{+,0}$ and $\Xi_{c}(2815)^{+,0}$ $\leftrightarrow$ $\Xi_{c}$ $1P(\frac{1}{2}^{-},\frac{3}{2}^{-})$
 as shown in Tables I and II. The measured masses can be well reproduced in our calculations and the deviation is usually less than 14 MeV. Additionally, it needs to be noted that $\Xi_{c}(2645)$ in the PDG should be labelled with $\Xi_{c}^{'}(2645)$.

 $\Xi_{c}(2970)$, earlier known as $\Xi_{c}(2980)$, was first observed by the Belle~\cite{art30} in 2006. Now the quantum numbers of $\Xi_{c}(2970)$ are determined as $\frac{1}{2}^{+}$ in the latest PDG. In our calculations, the only candidate is the $2S(\frac{1}{2}^{+})$ of $\Xi_{c}$ as shown in Tables I. And the predicted mass is 15 MeV less than the experimental data. At last,
 $\Xi_{c}$(3055) and $\Xi_{c}$(3080) were observed by the BABAR~\cite{art31} and the Belle~\cite{art89,art30,art601}. As shown in the PDG, their spin and parity values have not yet been clear so far. According to the measured masses, $\Xi_{c}$(3055) and $\Xi_{c}$(3080) are likely to be the
 $1D$ doublet ($\frac{3}{2}^{+}$, $\frac{5}{2}^{+}$) of $\Xi_{c}$ in Table I or the $2S$ doublet ($\frac{1}{2}^{+}$, $\frac{3}{2}^{+}$) of $\Xi_{c}^{'}$ in Table II. Considering the system with a smaller root mean square radius (especially for $\langle r_{\lambda}^{2}\rangle ^{1/2}$) is more stable, the $2S$ doublet states should then be the ideal candidates.

Outside of the PDG data, the Belle and the LHCb observed four charm-strange baryons, namely $\Xi_{c}(2930)$~\cite{art120}, $\Xi_{c}(2923)$, $\Xi_{c}(2939)$, and $\Xi_{c}(2964)$~\cite{art125}, whose masses are very close to each other. By the predicted masses in Tables I and II, the above four baryons can be assigned to be the first orbital ($1P$) excitation of $\Xi_{c}^{'}$ or the first radial (2S) excitation of $\Xi_{c}$. At present, we can not determine their quantum numbers accurately.

The $\Xi_{c}$(3123) was observed by the BABAR Collaboration~\cite{art31}. However, it has not yet appeared in the PDG~\cite{art2} so far. In our calculations, the predicted masses of the $\Xi_{c}$ $3S(\frac{1}{2}^{+})$ and  $\Xi_{c}^{'}$ $2S(\frac{3}{2}^{+})$ are 3155 MeV and 3095 MeV, respectively, which are relatively closer to the measured value of the $\Xi_{c}$(3123) than those of other states. Because the $\Xi_{c}^{'}$ $2S(\frac{3}{2}^{+})$ has been considered as the candidate of the $\Xi_{c}(3080)$, the $\Xi_{c}$(3123) is likely to be the $\Xi_{c}$ $3S(\frac{1}{2}^{+})$ state. Of course, whether $\Xi_{c}$(3123) exists remains to be tested.

 The calculated mass spectra of $\Xi_{b}$ and $\Xi^{'}_{b}$ families are listed in Tables IV-VI where a total of six bottom-strange baryons with determined quantum numbers in the PDG have been assigned to the possible states, such as $\Xi_{b}^{-,0}$ $\leftrightarrow$ $\Xi_{b}$ $1S(\frac{1}{2}^{+})$, $\Xi_{b}^{'}(5935)^{-}$ $\leftrightarrow$ $\Xi_{b}^{'}$ $1S(\frac{1}{2}^{+})$, $\Xi_{b}(5945)^{0}$ and $\Xi_{b}(5955)^{-}$ $\leftrightarrow$ $\Xi_{b}^{'}$ $1S(\frac{3}{2}^{+})$.

In 2021, AMS collaboration determined the $\Xi_{b}(6100)$ with the quantum numbers $J^{P}=\frac{3}{2}^{-}$ by measuring the typical decay chain of $\Xi_{b}(6100)^{-}\rightarrow \Xi_{b}^{*0}\pi^{-}\rightarrow\Xi_{b}^{-}\pi^{+}\pi^{-}$~\cite{art90}. Very recently, the values of $J^{P}=\frac{3}{2}^{-}$ of $\Xi_{b}(6100)$ were written into the PDG data. Table IV shows the mass of the $\Xi_{b}(6100)$ is very close to that of the $\Xi_{b}$ $1P(\frac{3}{2}^{-})$ state. So, the $\Xi_{b}(6100)$ is most likely to be the $1P(\frac{3}{2}^{-})$ state of $\Xi_{b}$. From Tables IV and V, we find the experimental data can be well reproduced by our theoretical calculations. In addition, it should be pointed out $\Xi_{b}(5945)^{0}$ and $\Xi_{b}(5955)^{-}$ in the PDG ought to be labelled with $\Xi_{b}^{'}(5945)^{0}$ and $\Xi_{b}^{'}(5955)^{-}$.

The last two $\Xi_{b}$ baryons in the PDG, $\Xi_{b}(6227)^{-}$ and $\Xi_{b}(6227)^{0}$, were reported by the LHCb Collaboration in 2018~\cite{art130}. But their spin and parity values are still not confirmed. In Tables IV and V, there are six states (one $2S$ state of $\Xi_{b}$ and five $1P$ states of $\Xi_{b}^{'}$) whose masses range from 6224 MeV to 6243 MeV. Each of them could be considered as a possible assignment to the $\Xi_{b}(6227)$.

In 2021, two bottom-strange baryons $\Xi_{b}(6327)$ and $\Xi_{b}(6333)$ were reported by the LHCb Collaboration. Very recently, the LHCb implied in experiment that they should belong to the $\Xi_{b}$ 1D($\frac{3}{2}^{+}$, $\frac{5}{2}^{+}$) doublet~\cite{art126}. In Table IV, one can see the predicted masses of the 1D doublet ($\frac{3}{2}^{+}$, $\frac{5}{2}^{+}$) of $\Xi_{b}$ indeed match with the experimental data of the $\Xi_{b}(6327)$ and $\Xi_{b}(6333)$.

\subsection*{3.5 Shell structure of the mass spectra }
 The shell structure of the mass spectra are presented in Figs.10 and 11, where
 only the states with their root mean square radii less than 1 fm are collected. In this two figures, the well established baryons in experiment are labeled next to the corresponding states, and the recently observed baryons in experiment are arranged in their possible positions preliminarily according to the discussions in subsection 3.4.

  From these two figures, we could get a bird's-eye view of the mass spectra. Firstly, the baryon spectra of the $\Xi_{c}$ ($\Xi_{c}^{'}$) and $\Xi_{b}$ ($\Xi_{b}^{'}$) almost have the same shell structure. Secondly, the baryons with lighter masses were discovered earlier in experiment.
    Thirdly, there is a serious energy degeneracy in the $P $, $D$ and $F$ states for the $\Xi_{c}^{'}$ ($\Xi_{b}^{'}$) family. And the calculated masses for the $2S$ states of $\Xi_{c}$ ($\Xi_{b}$) family are very close to those of the $1P$ states of the $\Xi_{c}^{'}$ ($\Xi_{b}^{'}$) family. This makes these states hard to be identified in experiment.

    At last, the states which are possibly observed in experiment can be predicted. For the $\Xi_{c}$ family, the $1D$ doublets and the $3S(1/2)^{+}$ are likely to be experimentally observed first. In the $\Xi_{c}^{'}$ family, the $3S$ doublets might be the next ones to be observed. However, the predicted mass range of $3S$ doublet states overlaps heavily with that of the $1D$ states.
    As to the $\Xi_{b}$ family, we find the $1P(1/2)^{-}$ state should have been discovered earlier in experiment and the predicted mass is $6084$ MeV. The possibly observed states in the $\Xi_{b}^{'}$ family should be the $1P$ states where experimental observations will encounter the difficulty of the energy degeneracy.

\begin{figure}[htbp]
\begin{center}
\includegraphics[width=0.8\textwidth]{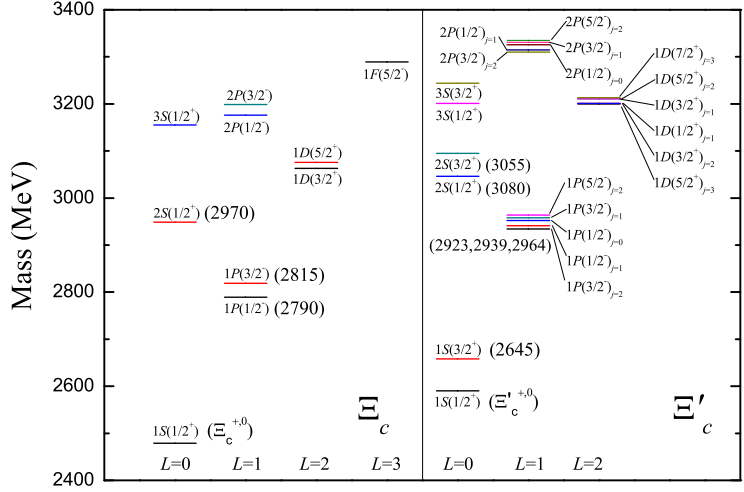}
\end{center}
\caption{(Color online)Mass spectrum shells of the $\Xi_{c}$ and $\Xi_{c}^{'}$ families.}
\end{figure}

\begin{figure}[htbp]
\begin{center}
\includegraphics[width=0.8\textwidth]{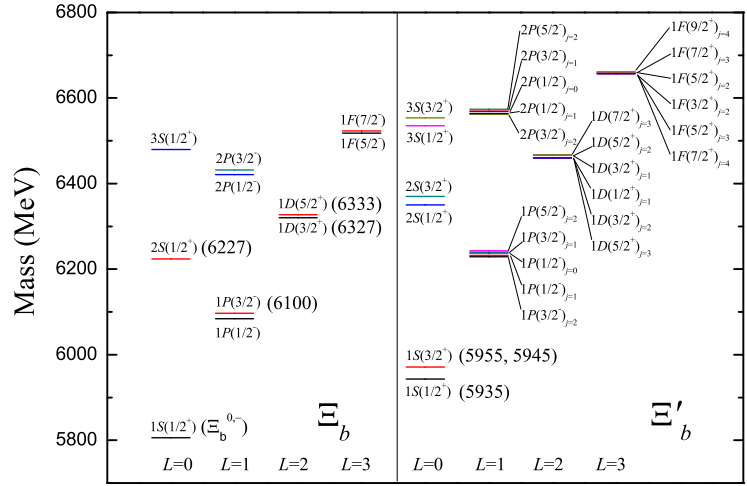}
\end{center}
\caption{(Color online)Mass spectrum shells of the $\Xi_{b}$ and $\Xi_{b}^{'}$ families.}
\end{figure}

\section*{IV. Conclusions}

Motivated by the experimental development of singly heavy baryons, we investigate the strange single heavy baryon spectra in a three-quark system, where the relativistic quark model and the ISG method are employed.  Considering the feature that the $\lambda$-mode appears lower in energy for definite states $nL(J^{P})$, we only focus on the $\lambda$-mode and obtain the mass spectra of the $\Xi_{c}$, $\Xi_{c}^{'}$, $\Xi_{b}$ and $\Xi_{b}^{'}$ families.
For the well established baryons, our predicted masses can nicely reproduce the experimental data. We also investigate the root mean square radii and the radial probability density distributions, from which we can learn more about the structure of the strange single heavy baryons.

Based on the predicted mass spectra, we construct the Regge trajectories in the $(J,M^{2})$ plane. Nevertheless, we can not currently construct the linear trajectories in the $(n,M^{2})$ plane, which is an apparent difference between our mass spectra and those in the relativistic quark-diquark picture~\cite{art1}.

For some recently observed baryons, We have preliminarily determined their reasonable positions in the mass spectra.
At last, the mass spectral structure of the $\Xi_{c}$ ($\Xi_{c}^{'}$) and $\Xi_{b}$ ($\Xi_{b}^{'}$) families is presented, from which we could get a bird's-eye view of the mass spectra and easily foresee where the experiment is going. Then, we analyze some states which might be observed in the forthcoming experiments.

\section*{Acknowledgements}

Zhen-yu Li, one of the authors, thanks Wen-chao Dong for his valuable reference, helpful discussion and kind help in programming. And Li is also grateful to Professor Xian-Jian Shi for his support and encouragement. This work could not have been done without the joint efforts of all members of the phenomenological QCD theory group in North China Electric Power University. This research is supported by the Science and Technology Talent Project of Education Bureau of Guizhou Province, China (QJHKY [2018]058)), the National Natural Science Foundation of China (Grant No. 11675265), the Continuous Basic Scientific Research Project (Grant No. WDJC-2019-13), and the Leading Innovation Project (Grant No. LC 192209000701).

\section*{Appendix}

\begin{table*}[htbp]
\begin{ruledtabular}\caption{The root mean square radius (fm) and the mass spectrum (MeV) of the $\Xi_{c}$ family.}
\begin{tabular}{c c c c c c c c c}
$l_{\rho}$  $l_{\lambda}$ $L$ $s$ $j$  &$nL$($J^{P}$) & $\langle r_{\rho}^{2}\rangle^{1/2}$ & $\langle r_{\lambda}^{2}\rangle^{1/2}$ & mass & exp.\cite{art2} &\cite{art1} \\ \hline
\multirow{4}{*}{0 0 0 0 0}
         & $1S$($\frac{1}{2}^{+}$) & 0.512 & 0.437 & 2479 &\tabincell{c}{ $\Xi_{c}^{+}$ 2467.71(0.23)\\$\Xi_{c}^{0}$ 2470.44(0.28)} & 2476  \\
         & $2S$($\frac{1}{2}^{+}$) & 0.645 & 0.768 & 2949 & 2970? & 2959  \\
         & $3S$($\frac{1}{2}^{+}$) & 0.968 & 0.607 & 3155 & 3123?\cite{art31} & 3323  \\
         & $4S$($\frac{1}{2}^{+}$) & 0.690 & 1.131 & 3318 & ~ & 3632 \\ \hline
\multirow{4}{*}{0 1 1 0 1}
         & $1P$($\frac{1}{2}^{-}$) & 0.542 & 0.627 & 2789 &\tabincell{c}{ 2791.9(0.5)\\ 2793.9(0.5)} & 2792  \\
         & $2P$($\frac{1}{2}^{-}$) & 0.615 & 0.948 & 3176 & ~ & 3179  \\
        & $3P$($\frac{1}{2}^{-}$)  & 1.038 & 0.763 & 3390 & ~ & 3500 \\
         & $4P$($\frac{1}{2}^{-}$) & 0.655 & 1.285 & 3492 & ~ & 3785  \\ \hline
\multirow{4}{*}{0 1 1 0 1}
         & $1P$($\frac{3}{2}^{-}$) & 0.550 & 0.654 & 2819 &\tabincell{c}{ 2816.51(0.25)\\2819.79(0.30)} & 2819  \\
         & $2P$($\frac{3}{2}^{-}$) & 0.613 & 0.977 & 3199 & ~ & 3201  \\
         & $3P$($\frac{3}{2}^{-}$) & 1.053 & 0.779 & 3412 & ~ & 3519  \\
         & $4P$($\frac{3}{2}^{-}$) & 0.645 & 1.278 & 3508 & ~ & 3804  \\ \hline
\multirow{4}{*}{0 2 2 0 2}
         & $1D$($\frac{3}{2}^{+}$) & 0.561 & 0.825 & 3063 & 3055? & 3059  \\
         & $2D$($\frac{3}{2}^{+}$) & 0.601 & 1.161 & 3406 & ~ & 3388\\
         & $3D$($\frac{3}{2}^{+}$) & 1.084 & 0.936 & 3617 & ~ & 3678\\
         & $4D$($\frac{3}{2}^{+}$) & 0.627 & 1.349 & 3676 & ~ & 3945 \\ \hline
\multirow{4}{*}{0 2 2 0 2}
         & $1D$($\frac{5}{2}^{+}$) & 0.565 & 0.843 & 3076 & 3080? & 3076 \\
         & $2D$($\frac{5}{2}^{+}$) & 0.604 & 1.190 & 3419 & ~ & 3407   \\
         & $3D$($\frac{5}{2}^{+}$) & 1.092 & 0.945 & 3627 & ~ & 3699  \\
         & $4D$($\frac{5}{2}^{+}$) & 0.618 & 1.328 & 3688 & ~ & 3965  \\ \hline
\multirow{4}{*}{0 3 3 0 3}
         & $1F$($\frac{5}{2}^{-}$) & 0.567 & 0.998 & 3289 & ~ & 3278 \\
         & $2F$($\frac{5}{2}^{-}$) & 0.604 & 1.413 & 3613 & ~ & 3572 \\
         & $3F$($\frac{5}{2}^{-}$) & 1.103 & 1.087 & 3817 & ~ & 3845 \\
         & $4F$($\frac{5}{2}^{-}$) & 0.602 & 1.314 & 3861 & ~ & 4098  \\ \hline
\multirow{4}{*}{0 3 3 0 3}
         & $1F$($\frac{7}{2}^{-}$) & 0.569 & 1.009 & 3294 & ~ & 3292  \\
         & $2F$($\frac{7}{2}^{-}$) & 0.607 & 1.439 & 3619 & ~ & 3592  \\
         & $3F$($\frac{7}{2}^{-}$) & 1.111 & 1.088 & 3821 & ~ & 3865  \\
         & $4F$($\frac{7}{2}^{-}$) & 0.589 & 1.290 & 3871 & ~ & 4120  \\ \hline
\multirow{4}{*}{0 4 4 0 4}
         & $1G$($\frac{7}{2}^{+}$) & 0.566 & 1.147 & 3486 & ~ & 3469  \\
         & $2G$($\frac{7}{2}^{+}$) & 0.612 & 1.676 & 3798 & ~ & 3745 \\
         & $3G$($\frac{7}{2}^{+}$) & 1.121 & 1.205 & 4000 & ~ & ~ \\
         & $4G$($\frac{7}{2}^{+}$) & 0.559 & 1.222 & 4054 & ~ & ~  \\ \hline
\multirow{4}{*}{0 4 4 0 4}
         & $1G$($\frac{9}{2}^{+}$) & 0.567 & 1.154 & 3487 & ~ & 3483 \\
         & $2G$($\frac{9}{2}^{+}$) & 0.613 & 1.692 & 3799 & ~ & 3763  \\
         & $3G$($\frac{9}{2}^{+}$) & 1.126 & 1.208 & 4001 & ~ & ~ \\
         & $4G$($\frac{9}{2}^{+}$) & 0.551 & 1.202 & 4064 & ~ & ~  \\
\end{tabular}
\end{ruledtabular}
\end{table*}

\begin{table*}[htbp]
\begin{ruledtabular}\caption{The root mean square radius (fm) and the mass spectrum (MeV) of the $\Xi^{'}_{c}$ family (Part I).}
\begin{tabular}{c c c c c c c c c c}
$l_{\rho}$  $l_{\lambda}$ $L$ $s$ $j$  &$nL$($J^{P}$) & $\langle r_{\rho}^{2}\rangle^{1/2}$ & $\langle r_{\lambda}^{2}\rangle^{1/2}$ & mass &exp.\cite{art2} &\cite{art1} \\ \hline
\multirow{4}{*}{0 0 0 1 1 }
        & $1S$($\frac{1}{2}^{+}$)  & 0.590 & 0.431 & 2590 & \tabincell{c}{$\Xi_{c}^{'+}$ 2578.2(0.5)\\$\Xi_{c}^{'0}$ 2578.7(0.5)} & 2579   \\
        & $2S$($\frac{1}{2}^{+}$)  & 0.821 & 0.705& 3046 & 3055? & 2983   \\
        & $3S$($\frac{1}{2}^{+}$)  & 0.918 & 0.671& 3201 & ~ & 3377  \\
        & $4S$($\frac{1}{2}^{+}$) & 0.897 & 1.053 & 3425 & ~ & 3695   \\ \hline
\multirow{4}{*}{0 0 0 1 1}
        & $1S$($\frac{3}{2}^{+}$)  & 0.611 & 0.476& 2658 & \tabincell{c}{2645.10(0.30)\\2646.16(0.25)} & 2654  \\
        & $2S$($\frac{3}{2}^{+}$)  & 0.801 & 0.763& 3095 & 3080? & 3026 \\
        & $3S$($\frac{3}{2}^{+}$)  & 0.968 & 0.676& 3244 & ~ & 3396  \\
        & $4S$($\frac{3}{2}^{+}$)  & 0.850 & 1.095& 3456 & ~ & 3709   \\ \hline
\multirow{4}{*}{0 1 1 1 0}
        & $1P$($\frac{1}{2}^{-}$)  & 0.649 & 0.671& 2952 & ~ & 2936   \\
        & $2P$($\frac{1}{2}^{-}$)  & 0.762 & 0.978& 3326 & ~ & 3313   \\
        & $3P$($\frac{1}{2}^{-}$)  & 1.055 & 0.811& 3469 & ~ & 3630   \\
        & $4P$($\frac{1}{2}^{-}$)  & 0.783 & 1.245& 3636 & ~ & 3912   \\ \hline
\multirow{4}{*}{0 1 1 1 1}
        & $1P$($\frac{1}{2}^{-}$)  & 0.644 & 0.659& 2941 & ~ & 2854  \\
        & $2P$($\frac{1}{2}^{-}$)  & 0.763 & 0.962& 3315 & ~ & 3267   \\
        & $3P$($\frac{1}{2}^{-}$)  & 1.048 & 0.804& 3460 & ~ & 3598   \\
        & $4P$($\frac{1}{2}^{-}$)  & 0.788 & 1.248& 3628 & ~ & 3887   \\ \hline
\multirow{4}{*}{0 1 1 1 1}
        & $1P$($\frac{3}{2}^{-}$)  & 0.651 & 0.677& 2958 & ~ & 2935   \\
        & $2P$($\frac{3}{2}^{-}$)  & 0.761 & 0.985& 3331 & ~ & 3311   \\
        & $3P$($\frac{3}{2}^{-}$)  & 1.059 & 0.814& 3473 & ~ & 3628   \\
        & $4P$($\frac{3}{2}^{-}$)  & 0.780 & 1.243& 3640 & ~ & 3911   \\ \hline
\multirow{4}{*}{0 1 1 1 2}
        & $1P$($\frac{3}{2}^{-}$)  & 0.642 & 0.653& 2934 & 2930?\cite{art120} & 2912  \\
        & $2P$($\frac{3}{2}^{-}$)  & 0.764 & 0.955& 3310 & ~ & 3293  \\
        & $3P$($\frac{3}{2}^{-}$)  & 1.043 & 0.801& 3456 & ~ & 3613   \\
        & $4P$($\frac{3}{2}^{-}$)  & 0.791 & 1.249& 3624 & ~ & 3898   \\ \hline
\multirow{4}{*}{0 1 1 1 2}
        & $1P$($\frac{5}{2}^{-}$)  & 0.652 & 0.682& 2964 &    & 2929   \\
        & $2P$($\frac{5}{2}^{-}$) & 0.761 & 0.993 & 3335 & ~ & 3303   \\
        & $3P$($\frac{5}{2}^{-}$)  & 1.062 & 0.817& 3477 & ~ & 3619   \\
        & $4P$($\frac{5}{2}^{-}$)  & 0.778 & 1.241& 3644 & ~ & 3902   \\ \hline
\multirow{4}{*}{0 2 2 1 1}
        & $1D$($\frac{1}{2}^{+}$)  & 0.668 & 0.851& 3201 & ~ & 3163   \\
        & $2D$($\frac{1}{2}^{+}$)  & 0.744 & 1.195& 3541 & ~ & 3505   \\
        & $3D$($\frac{1}{2}^{+}$) & 1.104 & 0.955 & 3676 & ~ & ~   \\
        & $4D$($\frac{1}{2}^{+}$)  & 0.738 & 1.303& 3816 & ~ & ~   \\ \hline
\multirow{4}{*}{0 2 2 1 1}
        & $1D$($\frac{3}{2}^{+}$)  & 0.671 & 0.865& 3211 & ~ & 3167   \\
        & $2D$($\frac{3}{2}^{+}$)  & 0.745 & 1.219& 3550 & ~ & 3506   \\
        & $3D$($\frac{3}{2}^{+}$)  & 1.111 & 0.963& 3684 & ~ & ~   \\
        & $4D$($\frac{3}{2}^{+}$)  & 0.734 & 1.285& 3827 & ~ & ~   \\
\end{tabular}
\end{ruledtabular}
\end{table*}

\begin{table*}[htbp]
\begin{ruledtabular}\caption{The root mean square radius (fm) and the mass spectrum (MeV) of the $\Xi^{'}_{c}$ family (Part II).}
\begin{tabular}{c c c c c c c c c c c c}
$l_{\rho}$  $l_{\lambda}$ $L$ $s$ $j$  &$nL$($J^{P}$) & $\langle r_{\rho}^{2}\rangle^{1/2}$ & $\langle r_{\lambda}^{2}\rangle^{1/2}$ & mass &\cite{art1} \\ \hline
 \multirow{4}{*}{0 2 2 1 2}
        & $1D$($\frac{3}{2}^{+}$) & 0.668 & 0.851 & 3201  & 3160    \\
        & $2D$($\frac{3}{2}^{+}$) & 0.744 & 1.195 & 3541  & 3497    \\
        & $3D$($\frac{3}{2}^{+}$) & 1.104 & 0.955 & 3676  & ~     \\
        & $4D$($\frac{3}{2}^{+}$) & 0.738 & 1.303 & 3816  & ~    \\ \hline
\multirow{4}{*}{0 2 2 1 2}
        & $1D$($\frac{5}{2}^{+}$) & 0.671 & 0.865 & 3211  & 3166   \\
        & $2D$($\frac{5}{2}^{+}$) & 0.745 & 1.219 & 3551  & 3504  \\
        & $3D$($\frac{5}{2}^{+}$) & 1.111 & 0.963 & 3685  & ~     \\
        & $4D$($\frac{5}{2}^{+}$) & 0.734 & 1.285 & 3828  & ~  \\ \hline
\multirow{4}{*}{0 2 2 1 3}
        & $1D$($\frac{5}{2}^{+}$) & 0.667 & 0.850 & 3200  & 3153   \\
        & $2D$($\frac{5}{2}^{+}$) & 0.744 & 1.193 & 3540  & 3493  \\
        & $3D$($\frac{5}{2}^{+}$) & 1.104 & 0.954 & 3676  & ~   \\
        & $4D$($\frac{5}{2}^{+}$) & 0.738 & 1.304 & 3815  & ~  \\ \hline
\multirow{4}{*}{0 2 2 1 3}
        & $1D$($\frac{7}{2}^{+}$) & 0.672 & 0.868 & 3213  & 3147  \\
        & $2D$($\frac{7}{2}^{+}$) & 0.746 & 1.224 & 3552  & 3486   \\
        & $3D$($\frac{7}{2}^{+}$) & 1.112 & 0.965 & 3686  & ~    \\
        & $4D$($\frac{7}{2}^{+}$) & 0.733 & 1.281 & 3829  & ~    \\ \hline
\multirow{4}{*}{0 3 3 1 2}
        & $1F$($\frac{3}{2}^{-}$) & 0.676 & 1.022 & 3424  & 3418  \\
        & $2F$($\frac{3}{2}^{-}$) & 0.742 & 1.454 & 3744  & ~  \\
        & $3F$($\frac{3}{2}^{-}$) & 1.136 & 1.096 & 3872  & ~   \\
        & $4F$($\frac{3}{2}^{-}$) & 0.699 & 1.262 & 4010  & ~   \\ \hline
\multirow{4}{*}{0 3 3 1 2}
        & $1F$($\frac{5}{2}^{-}$) & 0.678 & 1.031 & 3428  & 3408  \\
        & $2F$($\frac{5}{2}^{-}$) & 0.744 & 1.474 & 3748  & ~    \\
        & $3F$($\frac{5}{2}^{-}$) & 1.140 & 1.101 & 3876  & ~   \\
        & $4F$($\frac{5}{2}^{-}$) & 0.698 & 1.243 & 4020  & ~    \\ \hline
\multirow{4}{*}{0 3 3 1 3}
        & $1F$($\frac{5}{2}^{-}$) & 0.676 & 1.022 & 3424  & 3394  \\
        & $2F$($\frac{5}{2}^{-}$) & 0.742 & 1.454 & 3744  & ~    \\
        & $3F$($\frac{5}{2}^{-}$) & 1.136 & 1.096 & 3872  & ~   \\
        & $4F$($\frac{5}{2}^{-}$) & 0.699 & 1.263 & 4009  & ~   \\ \hline
\multirow{4}{*}{0 3 3 1 3}
        & $1F$($\frac{7}{2}^{-}$) & 0.678 & 1.031 & 3428  & 3393  \\
        & $2F$($\frac{7}{2}^{-}$) & 0.744 & 1.475 & 3748  & ~   \\
        & $3F$($\frac{7}{2}^{-}$) & 1.140 & 1.101 & 3876  & ~   \\
        & $4F$($\frac{7}{2}^{-}$) & 0.698 & 1.242 & 4021  & ~     \\ \hline
\multirow{4}{*}{0 3 3 1 4}
        & $1F$($\frac{7}{2}^{-}$) & 0.676 & 1.021 & 3423  & 3373  \\
        & $2F$($\frac{7}{2}^{-}$) & 0.742 & 1.453 & 3744  & ~    \\
        & $3F$($\frac{7}{2}^{-}$) & 1.136 & 1.095 & 3872  & ~   \\
        & $4F$($\frac{7}{2}^{-}$) & 0.699 & 1.264 & 4009  & ~   \\ \hline
\multirow{4}{*}{0 3 3 1 4}
        & $1F$($\frac{9}{2}^{-}$) & 0.678 & 1.032 & 3428  & 3357  \\
        & $2F$($\frac{9}{2}^{-}$) & 0.744 & 1.476 & 3749  & ~   \\
        & $3F$($\frac{9}{2}^{-}$) & 1.140 & 1.102 & 3876  & ~   \\
        & $4F$($\frac{9}{2}^{-}$) & 0.698 & 1.241 & 4021  & ~     \\
\end{tabular}
\end{ruledtabular}
\end{table*}

\begin{table*}[htbp]
\begin{ruledtabular}\caption{The root mean square radius (fm) and the mass spectrum (MeV) of the $\Xi_{b}$ family.}
\begin{tabular}{c c c c c c c c}
$l_{\rho}$  $l_{\lambda}$ $L$ $s$ $j$  &$nL$($J^{P}$) & $\langle r_{\rho}^{2}\rangle^{1/2}$ & $\langle r_{\lambda}^{2}\rangle^{1/2}$ & mass & exp.\cite{art2} &\cite{art1} \\ \hline
\multirow{4}{*}{0 0 0 0 0}
        & $1S$($\frac{1}{2}^{+}$) & 0.518 & 0.400 & 5806 &\tabincell{c}{$\Xi_{b}^{-}$ 5797.0(0.6)\\$\Xi_{b}^{0}$ 5791.9(0.5)}  & 5803  \\
       & $2S$($\frac{1}{2}^{+}$) & 0.607 & 0.705 & 6224 & ~   & 6266  \\
       & $3S$($\frac{1}{2}^{+}$) & 0.990 & 0.549 & 6480 & ~  & 6601 \\
       & $4S$($\frac{1}{2}^{+}$) & 0.672 & 1.066 & 6568 & ~  & 6913 \\ \hline
\multirow{4}{*}{0 1 1 0 1}
         & $1P$($\frac{1}{2}^{-}$) & 0.539 & 0.571 & 6084 & ~  & 6120 \\
        & $2P$($\frac{1}{2}^{-}$) & 0.586 & 0.844 & 6421 & ~  & 6496 \\
       & $3P$($\frac{1}{2}^{-}$) & 1.034 & 0.713 & 6690 & ~  & 6805\\
         & $4P$($\frac{1}{2}^{-}$) & 0.673 & 1.281 & 6732 & ~  & 7068  \\ \hline
\multirow{4}{*}{0 1 1 0 1}
        & $1P$($\frac{3}{2}^{-}$) & 0.543 & 0.583 & 6097 & 6100.3(0.6)  & 6130  \\
         & $2P$($\frac{3}{2}^{-}$) & 0.585 & 0.853 & 6432 & ~  & 6502 \\
         & $3P$($\frac{3}{2}^{-}$) & 1.043 & 0.719 & 6700 & ~  & 6810  \\
         & $4P$($\frac{3}{2}^{-}$) & 0.668 & 1.293 & 6739 & ~  & 7073  \\ \hline
\multirow{4}{*}{0 2 2 0 2}
        & $1D$($\frac{3}{2}^{+}$) & 0.551 & 0.743 & 6320 & ~  & 6366  \\
         & $2D$($\frac{3}{2}^{+}$) & 0.568 & 0.962 & 6613 & ~  & 6690 \\
         & $3D$($\frac{3}{2}^{+}$) & 0.990 & 1.040 & 6883 & ~  & 6966  \\
         & $4D$($\frac{3}{2}^{+}$) & 0.778 & 1.359 & 6890 & ~  & 7208 \\ \hline
\multirow{4}{*}{0 2 2 0 2}
         & $1D$($\frac{5}{2}^{+}$) & 0.553 & 0.751 & 6327 & ~  & 6373 \\
         & $2D$($\frac{5}{2}^{+}$) & 0.568 & 0.967 & 6621 & ~  & 6696  \\
        & $3D$($\frac{5}{2}^{+}$) & 0.948 & 1.124 & 6888 & ~  & 6970 \\
         & $4D$($\frac{5}{2}^{+}$) & 0.831 & 1.294 & 6894 & ~  & 7212 \\ \hline
\multirow{4}{*}{0 3 3 0 3}
        & $1F$($\frac{5}{2}^{-}$) & 0.555 & 0.903 & 6518 & ~  & 6577  \\
        & $2F$($\frac{5}{2}^{-}$) & 0.553 & 1.064 & 6795 & ~  & 6863 \\
        & $3F$($\frac{5}{2}^{-}$) & 0.619 & 1.646 & 7032 & ~  & 7114 \\
         & $4F$($\frac{5}{2}^{-}$) & 1.110 & 0.960 & 7057 & ~  & 7339 \\ \hline
\multirow{4}{*}{0 3 3 0 3}
         & $1F$($\frac{7}{2}^{-}$) & 0.556 & 0.909 & 6523 & ~  & 6581 \\
       & $2F$($\frac{7}{2}^{-}$) & 0.553 & 1.070 & 6801 & ~  & 6867 \\
        & $3F$($\frac{7}{2}^{-}$) & 0.618 & 1.641 & 7034 & ~  & 7117  \\
        & $4F$($\frac{7}{2}^{-}$) & 1.111 & 0.965 & 7060 & ~  & 7342  \\ \hline
\multirow{4}{*}{0 4 4 0 4}
        & $1G$($\frac{7}{2}^{+}$) & 0.554 & 1.048 & 6692 & ~  & 6760  \\
         & $2G$($\frac{7}{2}^{+}$) & 0.542 & 1.178 & 6970 & ~  & 7020  \\
        & $3G$($\frac{7}{2}^{+}$) & 0.607 & 1.745 & 7167 & ~  & ~  \\
        & $4G$($\frac{7}{2}^{+}$) & 1.119 & 1.095 & 7214 & ~  & ~ \\ \hline
\multirow{4}{*}{0 4 4 0 4}
       & $1G$($\frac{9}{2}^{+}$) & 0.555 & 1.052 & 6695 & ~  & 6762 \\
        & $2G$($\frac{9}{2}^{+}$) & 0.544 & 1.189 & 6975 & ~  & 7032 \\
        & $3G$($\frac{9}{2}^{+}$) & 0.605 & 1.737 & 7169 & ~  & ~\\
        & $4G$($\frac{9}{2}^{+}$) & 1.120 & 1.098 & 7217 & ~  & ~  \\
\end{tabular}
\end{ruledtabular}
\end{table*}

\begin{table*}[htbp]
\begin{ruledtabular}\caption{The root mean square radius (fm) and the mass spectrum (MeV) of the $\Xi^{'}_{b}$ family (Part I).}
\begin{tabular}{c c c c c c c c c}
$l_{\rho}$  $l_{\lambda}$ $L$ $s$ $j$  &$nL$($J^{P}$) & $\langle r_{\rho}^{2}\rangle^{1/2}$ & $\langle r_{\lambda}^{2}\rangle^{1/2}$ & mass &exp.\cite{art2} &\cite{art1} \\ \hline
\multirow{4}{*}{0 0 0 1 1 }
       & $1S$($\frac{1}{2}^{+}$) & 0.604 & 0.411 & 5943 &5935.02(0.05)  & 5936  \\
       & $2S$($\frac{1}{2}^{+}$) & 0.741 & 0.697 & 6350 & ~  & 6329  \\
        & $3S$($\frac{1}{2}^{+}$) & 0.998 & 0.559 & 6535 & ~  & 6687 \\
       & $4S$($\frac{1}{2}^{+}$) & 0.804 & 1.063 & 6691 & ~  & 6978  \\ \hline
\multirow{4}{*}{0 0 0 1 1}
         & $1S$($\frac{3}{2}^{+}$) & 0.614 & 0.431 & 5971 & \tabincell{c}{5952.3(0.6)\\5955.33(0.13)}  & 5963 \\
        & $2S$($\frac{3}{2}^{+}$) & 0.735 & 0.716 & 6370 & ~  & 6342  \\
         & $3S$($\frac{3}{2}^{+}$) & 1.017 & 0.566 & 6554 & ~  & 6695  \\
         & $4S$($\frac{3}{2}^{+}$) & 0.793 & 1.087 & 6705 & ~  & 6984  \\ \hline
\multirow{4}{*}{0 1 1 1 0}
         & $1P$($\frac{1}{2}^{-}$) & 0.642 & 0.608 & 6238 & ~  & 6233  \\
         & $2P$($\frac{1}{2}^{-}$) & 0.709 & 0.866 & 6569 & ~  & 6611  \\
        & $3P$($\frac{1}{2}^{-}$) & 1.074 & 0.705 & 6758 & ~  & 6915  \\
         & $4P$($\frac{1}{2}^{-}$) & 0.772 & 1.296 & 6866 & ~  & 7174  \\ \hline
\multirow{4}{*}{0 1 1 1 1}
         & $1P$($\frac{1}{2}^{-}$) & 0.640 & 0.603 & 6232 & ~  & 6227  \\
         & $2P$($\frac{1}{2}^{-}$) & 0.709 & 0.862 & 6564 & ~  & 6604  \\
        & $3P$($\frac{1}{2}^{-}$) & 1.071 & 0.701 & 6754 & ~  & 6904  \\
         & $4P$($\frac{1}{2}^{-}$) & 0.774 & 1.292 & 6863 & ~  & 7164 \\ \hline
\multirow{4}{*}{0 1 1 1 1}
         & $1P$($\frac{3}{2}^{-}$) & 0.643 & 0.610 & 6240 & ~  & 6234 \\
         & $2P$($\frac{3}{2}^{-}$) & 0.709 & 0.868 & 6572 & ~  & 6605  \\
         & $3P$($\frac{3}{2}^{-}$) & 1.076 & 0.707 & 6760 & ~  & 6905  \\
         & $4P$($\frac{3}{2}^{-}$) & 0.771 & 1.297 & 6868 & ~  & 7163  \\ \hline
\multirow{4}{*}{0 1 1 1 2}
         & $1P$($\frac{3}{2}^{-}$) & 0.639 & 0.600 & 6229 & \tabincell{c}{6227.9(0.9)?\\6226.8(1.6)?}  & 6224  \\
        & $2P$($\frac{3}{2}^{-}$) & 0.709 & 0.859 & 6562 & ~  & 6598  \\
         & $3P$($\frac{3}{2}^{-}$) & 1.069 & 0.700 & 6752 & ~  & 6900  \\
        & $4P$($\frac{3}{2}^{-}$) & 0.774 & 1.291 & 6861 & ~  & 7159  \\ \hline
\multirow{4}{*}{0 1 1 1 2}
        & $1P$($\frac{5}{2}^{-}$) & 0.644 & 0.613 & 6243 & ~  & 6226 \\
         & $2P$($\frac{5}{2}^{-}$) & 0.708 & 0.871 & 6574 & ~  & 6596  \\
        & $3P$($\frac{5}{2}^{-}$) & 1.077 & 0.708 & 6762 & ~  & 6897  \\
       & $4P$($\frac{5}{2}^{-}$) & 0.770 & 1.299 & 6869 & ~  & 7156  \\ \hline
\multirow{4}{*}{0 2 2 1 1}
        & $1D$($\frac{1}{2}^{+}$) & 0.656 & 0.773 & 6460 & ~  & 6447  \\
        & $2D$($\frac{1}{2}^{+}$) & 0.690 & 0.992 & 6757 & ~  & 6767  \\
         & $3D$($\frac{1}{2}^{+}$) & 1.109 & 0.845 & 6941 & ~  & ~  \\
         & $4D$($\frac{1}{2}^{+}$) & 0.754 & 1.464 & 7017 & ~  & ~  \\ \hline
\multirow{4}{*}{0 2 2 1 1}
        & $1D$($\frac{3}{2}^{+}$) & 0.658 & 0.780 & 6466 & ~  & 6459  \\
        & $2D$($\frac{3}{2}^{+}$) & 0.690 & 0.998 & 6763 & ~  & 6775  \\
         & $3D$($\frac{3}{2}^{+}$) & 1.111 & 0.850 & 6946 & ~  & ~  \\
       & $4D$($\frac{3}{2}^{+}$) & 0.751 & 1.462 & 7020 & ~  & ~  \\

\end{tabular}
\end{ruledtabular}
\end{table*}

\begin{table*}[htbp]
\begin{ruledtabular}\caption{The root mean square radius (fm) and the mass spectrum (MeV) of the $\Xi^{'}_{b}$ family (Part II).}
\begin{tabular}{c c c c c c c c c}
$l_{\rho}$  $l_{\lambda}$ $L$ $s$ $j$  &$nL$($J^{P}$) & $\langle r_{\rho}^{2}\rangle^{1/2}$ & $\langle r_{\lambda}^{2}\rangle^{1/2}$ & mass &\cite{art1} \\ \hline
\multirow{4}{*}{0 2 2 1 2}
        & $1D$($\frac{3}{2}^{+}$) & 0.656 & 0.773 & 6460   & 6431  \\
        & $2D$($\frac{3}{2}^{+}$) & 0.690 & 0.992 & 6758  & 6751    \\
       & $3D$($\frac{3}{2}^{+}$) & 1.109 & 0.845 & 6941   & ~    \\
         & $4D$($\frac{3}{2}^{+}$) & 0.754 & 1.464 & 7017   & ~   \\ \hline
\multirow{4}{*}{0 2 2 1 2}
         & $1D$($\frac{5}{2}^{+}$) & 0.658 & 0.780 & 6466  & 6432  \\
         & $2D$($\frac{5}{2}^{+}$) & 0.690 & 0.999 & 6764   & 6751    \\
        & $3D$($\frac{5}{2}^{+}$) & 1.112 & 0.851 & 6946   & ~   \\
         & $4D$($\frac{5}{2}^{+}$) & 0.751 & 1.462 & 7021   & ~    \\ \hline
\multirow{4}{*}{0 2 2 1 3}
       & $1D$($\frac{5}{2}^{+}$) & 0.656 & 0.773 & 6460   & 6420  \\
         & $2D$($\frac{5}{2}^{+}$) & 0.690 & 0.991 & 6757   & 6740  \\
        & $3D$($\frac{5}{2}^{+}$) & 1.108 & 0.844 & 6941   & ~   \\
       & $4D$($\frac{5}{2}^{+}$) & 0.754 & 1.464 & 7017   & ~   \\ \hline
\multirow{4}{*}{0 2 2 1 3}
       & $1D$($\frac{7}{2}^{+}$) & 0.658 & 0.781 & 6467   & 6414  \\
        & $2D$($\frac{7}{2}^{+}$) & 0.690 & 1.000 & 6765   & 6736   \\
         & $3D$($\frac{7}{2}^{+}$) & 1.112 & 0.851 & 6946   & ~  \\
        & $4D$($\frac{7}{2}^{+}$) & 0.751 & 1.461 & 7021   & ~  \\ \hline
\multirow{4}{*}{0 3 3 1 2}
       & $1F$($\frac{3}{2}^{-}$) & 0.663 & 0.931 & 6657   & 6675   \\
        & $2F$($\frac{3}{2}^{-}$) & 0.678 & 1.121 & 6942   & ~  \\
         & $3F$($\frac{3}{2}^{-}$) & 1.130 & 0.986 & 7110   & ~  \\
         & $4F$($\frac{3}{2}^{-}$) & 0.734 & 1.580 & 7162   & ~  \\ \hline
\multirow{4}{*}{0 3 3 1 2}
         & $1F$($\frac{5}{2}^{-}$) & 0.664 & 0.936 & 6660   & 6686  \\
        & $2F$($\frac{5}{2}^{-}$) & 0.680 & 1.130 & 6946   & ~   \\
       & $3F$($\frac{5}{2}^{-}$) & 1.131 & 0.991 & 7114   & ~  \\
         & $4F$($\frac{5}{2}^{-}$) & 0.732 & 1.573 & 7164  & ~   \\ \hline
\multirow{4}{*}{0 3 3 1 3}
       & $1F$($\frac{5}{2}^{-}$) & 0.663 & 0.931 & 6657   & 6640  \\
        & $2F$($\frac{5}{2}^{-}$) & 0.678 & 1.121 & 6942   & ~  \\
         & $3F$($\frac{5}{2}^{-}$) & 1.130 & 0.986 & 7110   & ~  \\
        & $4F$($\frac{5}{2}^{-}$) & 0.734 & 1.580 & 7162   & ~ \\ \hline
\multirow{4}{*}{0 3 3 1 3}
         & $1F$($\frac{7}{2}^{-}$) & 0.664 & 0.936 & 6660   & 6641\\
        & $2F$($\frac{7}{2}^{-}$) & 0.680 & 1.131 & 6947   & ~ \\
       & $3F$($\frac{7}{2}^{-}$) & 1.131 & 0.991 & 7114   & ~ \\
        & $4F$($\frac{7}{2}^{-}$) & 0.731 & 1.572 & 7165   & ~\\ \hline
\multirow{4}{*}{0 3 3 1 4}
         & $1F$($\frac{7}{2}^{-}$) & 0.663 & 0.931 & 6657   & 6619\\
        & $2F$($\frac{7}{2}^{-}$) & 0.678 & 1.121 & 6942   & ~ \\
       & $3F$($\frac{7}{2}^{-}$) & 1.130 & 0.986 & 7110   & ~ \\
        & $4F$($\frac{7}{2}^{-}$) & 0.734 & 1.580 & 7162   & ~\\\hline
\multirow{4}{*}{0 3 3 1 4}
         & $1F$($\frac{9}{2}^{-}$) & 0.664 & 0.937 & 6661   & 6610\\
        & $2F$($\frac{9}{2}^{-}$) & 0.680 & 1.132 & 6947   & ~ \\
       & $3F$($\frac{9}{2}^{-}$) & 1.131 & 0.991 & 7114   & ~ \\
        & $4F$($\frac{9}{2}^{-}$) & 0.731 & 1.572 & 7165   & ~\\
\end{tabular}
\end{ruledtabular}
\end{table*}

\begin{table*}[htbp]
\begin{ruledtabular}\caption{Fitted values for the slope and intercept of the Regge trajectories for the $\Xi_{c}$ and $\Xi_{c}^{'}$ families.}
\begin{tabular}{c |c c| c c| c c c c c c c c c c c c c}
\multirow{2}{*}{Trajectory} & \multicolumn{2}{c|}{$n=1$}   &\multicolumn{2}{c|}{$n=2$} &\multicolumn{2}{c}{$n=3$}  \\
& \ $\alpha$(GeV$^{2}$) & \ $\beta$(GeV$^{2}$) & \ $\alpha$(GeV$^{2}$) & \ $\beta$(GeV$^{2}$) & \ $\alpha$(GeV$^{2}$) & \ $\beta$(GeV$^{2}$)   \\
\hline
$\bar{3}_{F}(NP)(NJ)$ &\ $1.493\pm0.056$ &\ $5.580\pm0.160$ &\ $ 1.433\pm0.022 $ &\ $8.047\pm0.063$ &\ $ 1.507\pm0.032 $ &\ $9.305\pm0.091$  \\
$\bar{3}_{F}(UP)(UJ)$  & \  $1.456\pm0.043$ & \ $7.122\pm0.098$  & \ $ 1.446\pm0.023$  & \ $9.399\pm0.052$ & \ $ 1.501\pm0.026$  & \ $10.784\pm0.059$  \\ \hline
$6_{F}(NP)(UJ)$ &\ $1.592\pm0.058$ &\ $6.098\pm0.166$ &\ $1.530\pm0.033$ &\ $8.611\pm0.096$ &\ $1.539\pm0.031$ &\ $9.574\pm0.089$  \\
$6_{F}(UP)(UJ) $  & \  $1.487\pm0.033$ & \ $7.959\pm0.076$  & \ $1.473\pm0.026$  & \ $10.292\pm0.059$ & \ $1.482\pm0.018$  & \ $11.260\pm0.041$  \\
$6_{F}(NP)(UJ) $  & \  $1.433\pm0.026$ & \ $9.545\pm0.044$  & \ $1.433\pm0.026$  & \ $11.837\pm0.045$ & \ $1.453\pm0.015$  & \ $12.795\pm0.026$  \\
$6_{F}(UP)(NJ) $  & \  $1.507\pm0.040$ & \ $4.933\pm0.153$  & \ $1.459\pm0.022$  & \ $7.451\pm0.082$ & \ $1.474\pm0.019$  & \ $8.371\pm0.071$  \\
$6_{F}(NP)(NJ) $  & \  $1.455\pm0.031$ & \ $6.618\pm0.098$  & \ $1.437\pm0.025$  & \ $8.980\pm0.079$ & \ $1.454\pm0.018$  & \ $9.911\pm0.058$  \\
$6_{F}(UP)(NJ) $  & \  $1.463\pm0.034$ & \ $8.041\pm0.078$  & \ $1.444\pm0.025$  & \ $10.383\pm0.057$ & \ $1.460\pm0.019$  & \ $11.336\pm0.043$  \\
\end{tabular}
\end{ruledtabular}
\end{table*}
\begin{table*}[htbp]
\begin{ruledtabular}\caption{Fitted values for the slope and intercept of the Regge trajectories for the $\Xi_{b}$ and $\Xi_{b}^{'}$ families.}
\begin{tabular}{c |c c| c c| c c c c c c c c c c c c c}
\multirow{2}{*}{Trajectory} & \multicolumn{2}{c|}{$n=1$}   &\multicolumn{2}{c|}{$n=2$} &\multicolumn{2}{c}{$n=3$}  \\
& \ $\alpha$(GeV$^{2}$) & \ $\beta$(GeV$^{2}$) & \ $\alpha$(GeV$^{2}$) & \ $\beta$(GeV$^{2}$) & \ $\alpha$(GeV$^{2}$) & \ $\beta$(GeV$^{2}$)   \\
\hline
$\bar{3}_{F}(NP)(NJ)$ &\ $2.760\pm0.134$ &\ $32.756\pm0.386$ &\ $2.470\pm0.027 $ &\ $37.595\pm0.077$ &\ $2.341\pm0.122$ &\ $41.187\pm0.350$  \\
$\bar{3}_{F}(UP)(UJ)$  & \  $2.582\pm0.099$ & \ $35.891\pm0.226$  & \ $2.449\pm0.014$  & \ $40.029\pm0.033$ & \ $ 2.190\pm0.114$  & \ $43.86\pm0.262$  \\\hline
$6_{F}(NP)(UJ)$ &\ $2.820\pm0.129$ &\ $34.316\pm0.371$ &\ $2.592\pm0.032$ &\ $39.116\pm0.092$ &\ $2.518\pm0.076$ &\ $41.681\pm0.219$  \\
$6_{F}(UP)(UJ) $  & \  $2.605\pm0.087$ & \ $37.677\pm0.199$  & \ $2.529\pm0.014$  & \ $41.849\pm0.033$ & \ $2.388\pm0.051$  & \ $44.509\pm0.116$  \\
$6_{F}(NP)(UJ) $  & \  $2.465\pm0.068$ & \ $40.538\pm0.117$  & \ $2.512\pm0.013$  & \ $44.409\pm0.022$ & \ $2.309\pm0.038$  & \ $47.045\pm0.065$  \\
$6_{F}(UP)(NJ) $  & \  $2.747\pm0.113$ & \ $31.888\pm0.428$  & \ $2.536\pm0.019$  & \ $36.834\pm0.072$ & \ $2.462\pm0.062$  & \ $39.454\pm0.235$  \\
$6_{F}(NP)(NJ) $  & \  $2.580\pm0.085$ & \ $35.207\pm0.273$  & \ $2.510\pm0.016$  & \ $39.450\pm0.050$ & \ $2.374\pm0.053$  & \ $42.222\pm0.170$  \\
$6_{F}(UP)(NJ) $  & \  $2.588\pm0.089$ & \ $37.766\pm0.205$  & \ $2.522\pm0.018$  & \ $41.921\pm0.041$ & \ $2.382\pm0.057$  & \ $44.573\pm0.131$  \\
\end{tabular}
\end{ruledtabular}
\end{table*}

\end{document}